\documentclass[12pt]{article}
\usepackage{appendix}
\usepackage{graphicx}
\usepackage{amsmath}
\usepackage{amsfonts}
\usepackage{color}    
\usepackage{bm}
\usepackage{subfigure}
\usepackage{epstopdf}
\usepackage{authblk}

\newcommand{\dwdg}{D_w / D_g}

\begin{document}

\title{Radiotherapy planning for glioblastoma based on a tumor growth model: \\Improving target volume delineation}

\author{Jan Unkelbach$^1$, Bjoern H. Menze$^{3,4}$, Ender Konukoglu$^2$, Florian Dittmann$^1$, Matthieu Le$^{1,3}$, Nicholas Ayache$^3$, Helen A. Shih$^1$}
\affil{$^1$ Department of Radiation Oncology, Massachusetts General Hospital and Harvard Medical School, Boston, MA, USA}
\affil{$^2$ Martino's Center for Biomedical Imaging, Massachusetts General Hospital and Harvard Medical School, Boston, MA, USA}
\affil{$^3$ Asclepios Project, INRIA Sophia Antipolis, France}
\affil{$^4$ Computer Vision Laboratory, ETH Z\"urich, Switzerland}


\maketitle

\begin{abstract}
Glioblastoma differ from many other tumors in the sense that they grow infiltratively into the brain tissue instead of forming a solid tumor mass with a defined boundary. Only the part of the tumor with high tumor cell density can be localized through imaging directly. In contrast, brain tissue infiltrated by tumor cells at low density appears normal on current imaging modalities. In current clinical practice, a uniform margin, typically two centimeters, is applied to account for microscopic spread of disease that is not directly assessable through imaging.

The current treatment planning procedure can potentially be improved by accounting for the anisotropy of tumor growth, which arises from different factors: Anatomical barriers such as the  falx cerebri represent boundaries for migrating tumor cells. In addition, tumor cells primarily spread in white matter and infiltrate gray matter at lower rate. We investigate the use of a phenomenological tumor growth model for treatment planning. The model is based on the Fisher-Kolmogorov equation, which formalizes these growth characteristics and estimates the spatial distribution of tumor cells in normal appearing regions of the brain. The target volume for radiotherapy planning can be defined as an isoline of the simulated tumor cell density.

This paper analyzes the model with respect to implications for target volume definition and identifies its most critical components. A retrospective study involving 10 glioblastoma patients treated at our institution has been performed. To illustrate the main findings of the study, a detailed case study is presented for a glioblastoma located close to the falx. In this situation, the falx represents a boundary for migrating tumor cells, whereas the corpus callosum provides a route for the tumor to spread to the contralateral hemisphere.  We further discuss the sensitivity of the model with respect to the input parameters. Correct segmentation of the brain appears to be the most crucial model input. We conclude that the tumor growth model provides a method to account for anisotropic growth patterns of glioma, and may therefore provide a tool to make target delineation more objective and automated.

\end{abstract}

\section{Introduction}
\label{SecIntro}
Gliomas are the most common primary brain tumors. For glioblastoma, the most aggressive form, median survival is a little more than one year after diagnosis, despite treatment schedules combining surgery, external beam radiotherapy, and chemotherapy. The clinical experience suggests that improvements in radiotherapy alone will not lead to patient cure. On the other hand, radiotherapy is the single most effective therapy. Hence, the benefit of radiotherapy to prolong survival is undoubted \cite{walker1979analysis,walker1978evaluation,kristiansen1981combined}. Consequently, we hypothesize that better targeting of the radiation can improve the efficacy of radiotherapy and lead to prolonged survival or reduced side effects.

Glioblastoma differ from many solid tumors in the sense that they grow infiltratively. Instead of forming a solid tumor mass with a defined boundary, glioblastoma are characterized by a smooth gradient of the tumor cell density. It is well known that tumor cells infiltrate the adjacent brain tissue and can be found several centimeters beyond the enhancing tumor mass that is visible on MRI \cite{Kelly1987,Watanabe1992,Price2006}. Functional imaging modalities including amino acid Positron Emission Tomography (PET), including FET (Fluoro-Ethyl-Tyrosine) and MET (Methionine) \cite{grosu2005methyl,miwa2004discrepancy,niyazi2011fet,goetz13}, have potential to improve the definition of the gross tumor volume. However, these modalities also fail to identify areas of low tumor cell infiltration.

In current practice, radiotherapy planning is typically based on the gross tumor volume (GTV) visible on MRI. Many practitioners account for the infiltrative growth by expanding the GTV with a 1-3 centimeter margin to form the clinical target volume (CTV), which is irradiated to a homogeneous dose of 60 Gy. The current treatment planning procedure can potentially be improved by accounting for two growth characteristics of gliomas that are currently not or not consistently incorporated in treatment planning:
\begin{itemize}
\item[$\bullet$ ] An anisotropic growth of gliomas. MRI data as well as histological analysis after autopsy or resection shows that glioma growth is anisotropic \cite{matsukado1961growth,coons99}. This accounts for the complex shapes of the visible tumor and is due to mainly three aspects:
\begin{itemize}
\item[-] Anatomical boundaries: The dura, including its extensions falx cerebri and tentorium cerebelli, represents a boundary for migrating tumor cells. Also, except for rare cases of CSF seeding, gliomas do not infiltrate the ventricles.
\item[-] Tumor cells infiltrate gray matter much less than white matter.
\item[-] Tumor cells seem to migrate primarily along white matter fiber tracts.  
\end{itemize}
\item[$\bullet$ ] A spatially varying tumor cell density. Most gliomas lack a defined boundary and tumor cells can be found several centimeters away from the T1 contrast enhancing tumor and outside of the peritumoral edema visible on T2. The tumor is rather characterized by a continuous fall-off of the tumor cell density \cite{Kelly1987,Watanabe1992}. 
\end{itemize}
These macroscopic growth characteristics are partly known from histopathological analysis after autopsy or resection. A comprehensive review can be found in \cite{coons99}.  
Mathematically, these macroscopic patterns of tumor evolution can be modeled by partial differential equations of reaction-diffusion type. In one of the most popular models, the Fisher-Kolmogorov model, tumor growth is described via two processes: local proliferation of tumor cells and migration into neighboring tissue \cite{harpold07}. The anisotropic growth due to anatomical boundaries and reduced gray matter infiltration is considered via a segmentation of the brain into white matter, gray matter and cerebrospinal fluid (CSF). Preferential spread of tumor cells along white matter fibers can be modeled based on a cell diffusion tensor derived from Diffusion Tensor Imaging (DTI) \cite{jbabdi05,Clatz2005}. This leads to a phenomenological tumor growth model that estimates the spatial distribution of tumor cells in the brain tissue in regions that appear normal using current imaging modalities. 
Incorporating the tumor growth model into radiotherapy planning can be approached in three steps:
\begin{itemize}
\item[$\bullet$ ] The shape of the target volume is modified as to match an isoline of the simulated tumor cell density while the total volume is kept identical to the conventional target based on an isotropic margin.
\item[$\bullet$ ] The prescription dose within the target is redistributed based on the varying tumor cell density, in order to deliver less dose to regions of low tumor cell density, and (possibly) boost the regions with high tumor cell density.
\item[$\bullet$ ] Unavoidable dose outside the target is pushed into regions that are more likely to be infiltrated by tumor cells.
\end{itemize}
This paper primarily addresses the first aspect, i.e. the use of the tumor growth model for automatic delineation of a radiotherapy target volume, taking into account spatial growth characteristics. An accompanying paper \cite{unkelbach13b} discusses the implications of the model for redistribution of dose, i.e. the second aspect. \\

\subsection{Previous work on the Fisher-Kolmogorov glioma model}
The Fisher-Kolmogorov model has been found to reproduce a wide range of glioma growth characteristics at a macroscopic scale. For a review regarding the development of the model see for example \cite{harpold07}. The mathematical properties of the Fisher-Kolmogorov equations are discussed in detail in \cite{murray02} and \cite{murray03}. A clinical application involves the personalization of the model to the patient specific anatomy (see e.g. \cite{Angelini2007} for a review). In addition, approaches to estimate of model parameters from imaging data have been proposed \cite{harpold07,konukoglu10a,Menze2011}, which are based on estimating the velocity of tumor growth and the steepness of the cell density fall-off. The velocity of tumor growth can be estimated from a sequence of MRI images \cite{konukoglu10a}; estimating the steepness of the tumor density fall-off is based on comparing the size of the contrast enhancing tumor core to the size of the peritumoral edema. These observable quantities relate to the generic model parameters proliferation rate and diffusion coefficient. The model has, in particular, been used in life time prediction \cite{Swanson2008,murray12}. In addition, several extensions of the model have been suggested. Those extensions integrate diffusion tensor imaging \cite{jbabdi05,Clatz2005}, tumor hypoxia \cite{gu11}, and the effect of radiotherapy \cite{rockne10} and chemotherapy on the evolution of the tumor cell density \cite{Tracqui1995,swanson02}. The idea of utilizing the model for target definition in radiotherapy has been suggested by several authors \cite{cobzas09,konukoglu10,bondiau2011biocomputing}, who propose ways of describing tumor infiltration beyond the visible margins of the tumor. Cobzas \cite{cobzas09} defines a distance metric in the brain based on DTI to describe the anisotropy in tumor growth. Konukoglu \cite{konukoglu10} and Bondiau \cite{bondiau2011biocomputing} uses a travelling wave approximation of the full reaction diffusion model to estimate the spatial distribution of tumor cells in normal appearing brain tissue.  


\subsection{Contributions and organization of this paper}
Even though the Fisher-Kolmogorov model has previously been studied in the mathematical biology and modeling community, an application for target delineation in clinical practice requires further research in multiple directions. Our work uses the previously published Fisher-Kolmogorov model and aims at characterizing the potential of the model for a true clinical application in radiotherapy treatment planning. We aim at developing a semi-automatic contouring tool for gliomas that is practically useful. The contributions of this paper are as follows:
\begin{itemize}
\item[$\bullet$] {\bf Systematic review of model assumptions:} In section \ref{SecModel} we provide a review of the tumor growth model in order to systematically compile the assumptions made to apply the model to target delineation. This includes 1) highlighting the qualitative features of the model that are relevant for radiotherapy planning, 2) identify the model parameters determining those features, and 3) illustrate how the imaging data is used to personalize the model for an application to the patient at hand. 
\item[$\bullet$] {\bf Use case characterization:} We provide a clear illustration of use cases of the tumor growth model for target delineation. In section \ref{SecCase} we perform a detailed case study for a tumor located in proximity to falx and corpus callosum, a situation where it is difficult to consistently account for anatomical boundaries in a manual delineation. In \ref{SecAppendix}, we show results for 4 additional cases with varying tumor location. The presented results illustrate the main findings from a study involving 10 glioblastoma patients previously treated at our institution.
\item[$\bullet$] {\bf Sensitivity analysis:} We discuss the sensitivity of the simulated tumor cell density with respect to the model inputs (section \ref{SecSensitivity}). In particular, we highlight the need for accurate brain segmentation into white matter, gray matter and anatomical barriers.
\item[$\bullet$] {\bf IMRT planning:} We compare radiotherapy plans optimized for intensity-modulated radiotherapy (IMRT) for model based versus manual target delineation (section \ref{SecPlanning}). We illustrate to what degree differences in target delineation are translated into differences in the delivered dose distribution.  
\end{itemize}
Furthermore, for one patient we compare the model predicted tumor cell density to the spatial growth patterns of the recurrent tumor (section \ref{SecValidation}). Finally, section \ref{SecDiscussion} discusses the scope of the work and the potential of a practical application of the tumor growth model for target delineation.



\section{Tumor growth calculations}
\label{SecModel}
The purpose of the tumor growth model amounts to estimating the density of tumor cells in the regions of the brain that appear normal on MRI. It is assumed that the growth of the tumor is described phenomenologically by local proliferation of tumor cells and migration into neighboring tissue. 

\begin{figure}[hbt]
\centering
\subfigure[]{
\includegraphics[height=6cm]{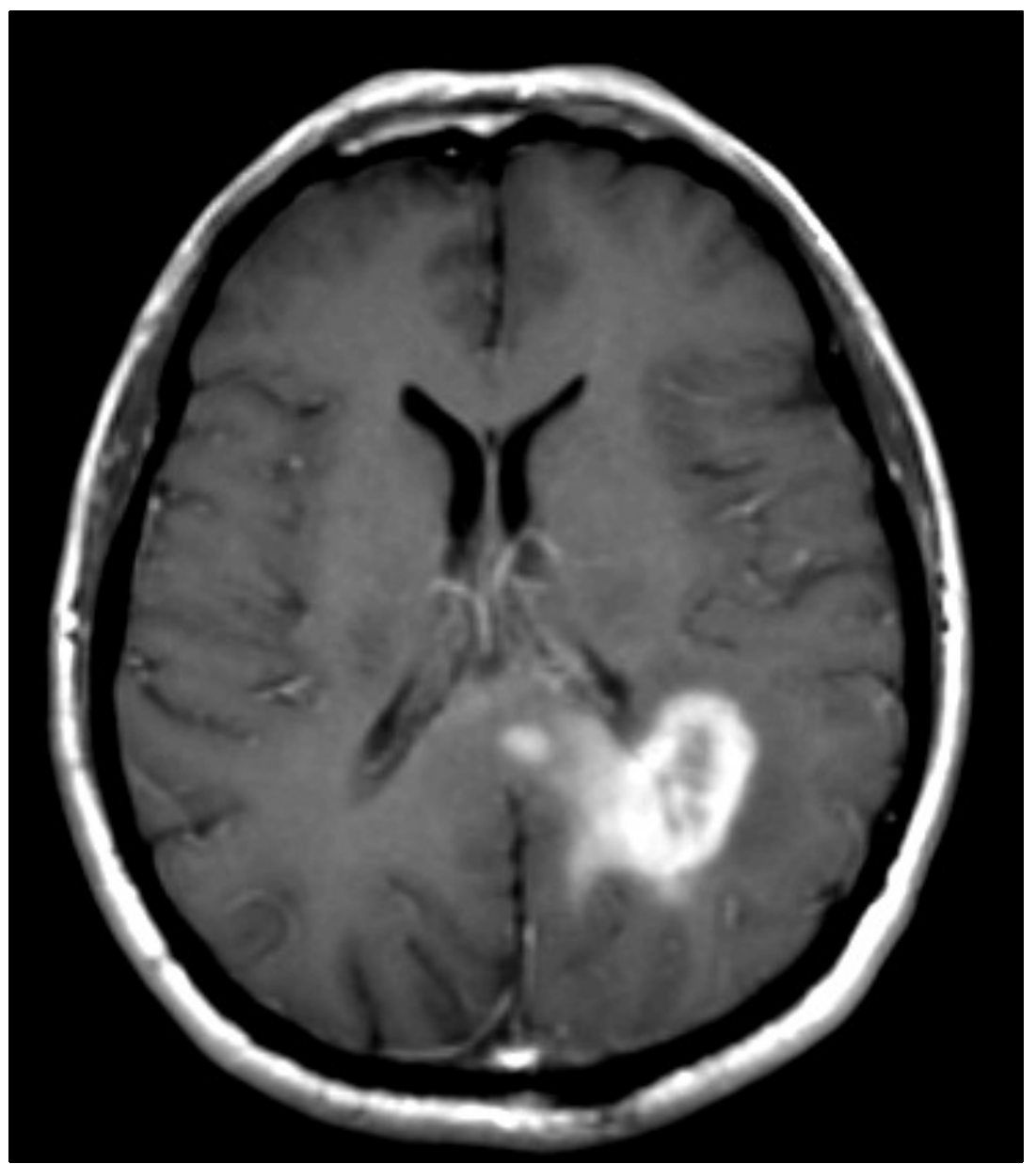}
}
\subfigure[]{
\includegraphics[height=6cm]{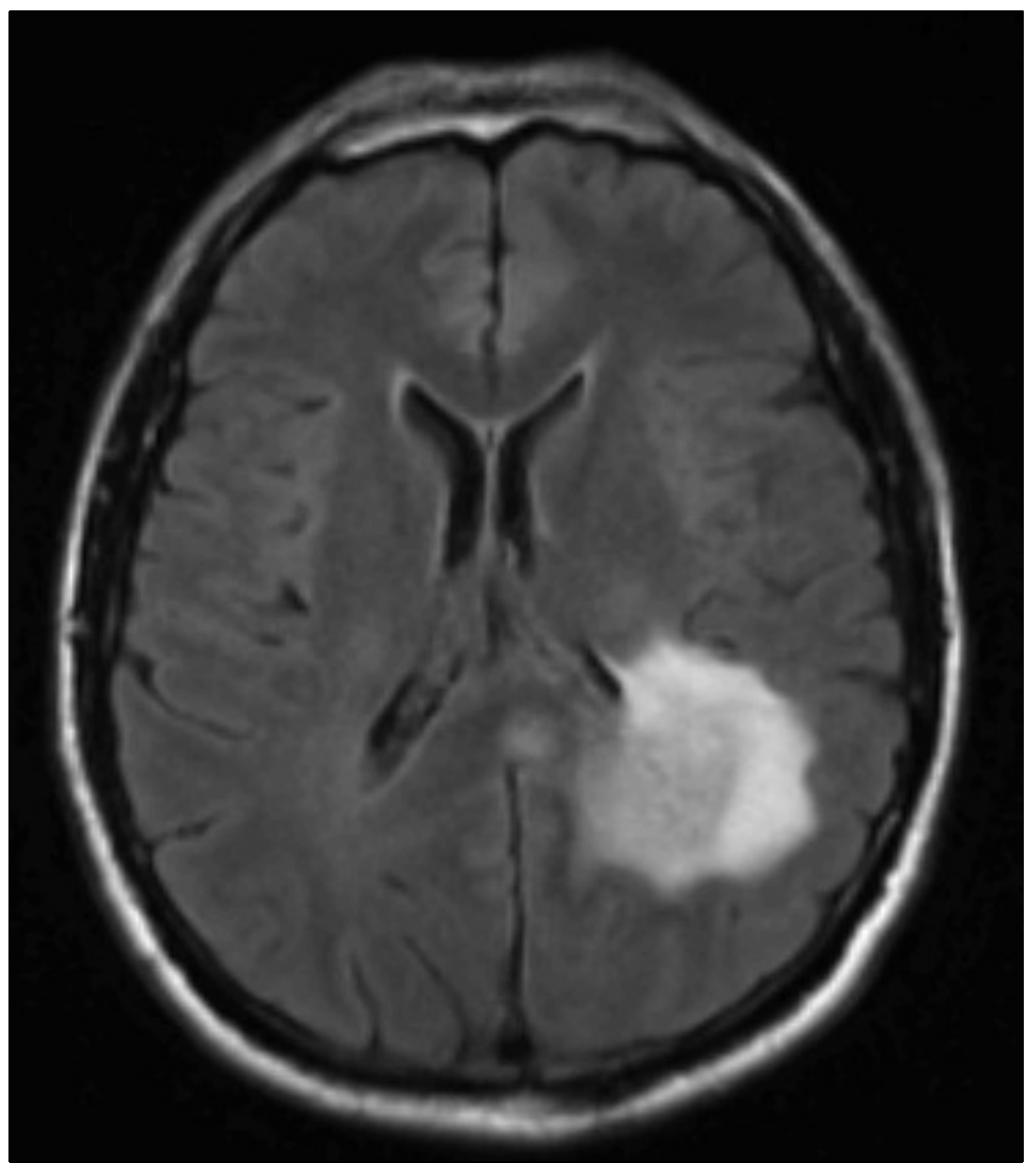}
}
\caption{(a) post-gadolinium T1 weighted image of a glioblastoma located in the left parietal lobe. (b) T2-FLAIR image of the same tumor, showing the surrounding peritumoral edema. (Note that the right side of the image corresponds to the left side of the brain to follow brain imaging conventions.)}
\label{FigMRI}
\end{figure}

\subsection{Patient specific input data}
The tumor growth calculations are based on MR images routinely acquired in clinical practice. The images incorporated into the simulation process are T1, T2, T2-FLAIR, and T1 post gadolinium. The T1 post gadolinium image shows the vascularized gross tumor volume, and the T2-FLAIR image shows the surrounding edematous region. For illustrative purpose, these two images are shown in figure \ref{FigMRI} for a case discussed in detail in this paper. 

\subsection{Data processing}
\begin{figure}[hbt]
\centering
\includegraphics[height=6cm]{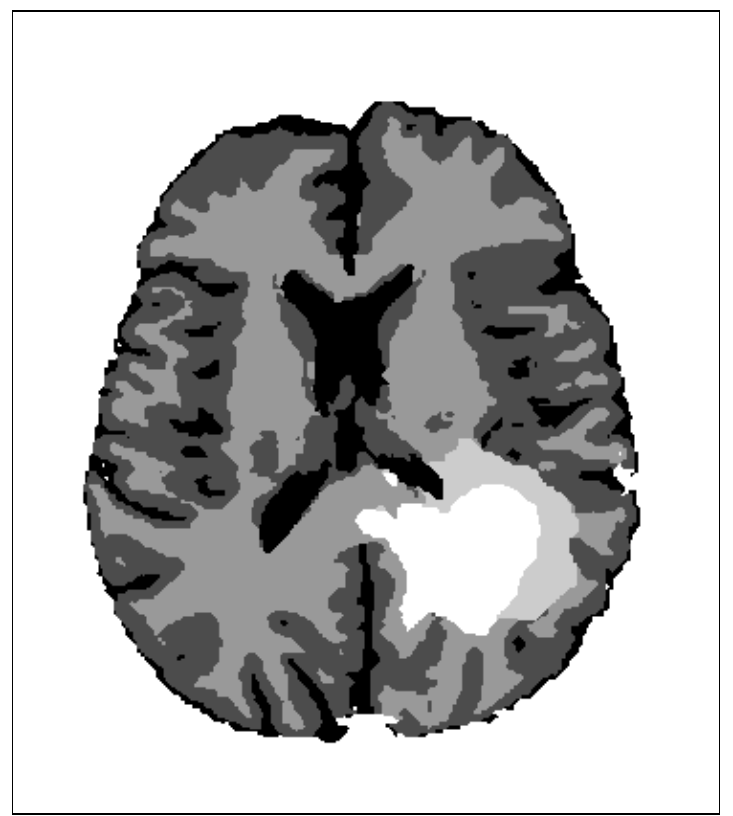}
\caption{Segmentation of the brain into contrast enhancing core (white), peritumoral edema, white matter, gray matter, and CSF (black).}
\label{FigSeg}
\end{figure}

In the first data processing step, all MR sequences are registered to the image with highest spatial resolution. This is done using rigid registration with 6 degrees of freedom, utilizing the function FLIRT \cite{jenkinson01} as part of the toolbox FSL \cite{smith04,woolrich09}. In the second step, a segmentation of the brain was obtained using the multimodal brain tumor segmentation algorithm published in \cite{menze10}. The algorithm is an Expectation-Maximization (EM) based segmentation method, which uses a probabilistic normal tissue atlas as spatial tissue prior. For every voxel, it estimates the posterior probability for three normal tissue classes (white matter, gray matter, and CSF\footnote{In this work, we refer to all brain tissue that is neither white matter, gray matter, nor tumor as CSF.}), as well as the lesion outlines on T1 post gadolinium and T2-FLAIR. The result of the EM segmentation is augmented as described in \cite{unkelbach12b} in order to facilitate a reliable identification of falx cerebri and tentorium cerebelli as anatomical barriers. The segmentation result for the patient in figure \ref{FigMRI} is shown in figure \ref{FigSeg}. In the last step, the reference MR image is registered to the radiotherapy planning CT using rigid registration. The transformation matrix is saved and later applied to register the simulated tumor cell density to the planning CT.

\subsection{Underlying tumor growth model}
\label{SecFKE}
It is assumed that tumor growth is described by two processes: local proliferation of tumor cells and migration of cells into neighboring brain tissue. Mathematically, this is formalized via the Fisher-Kolmogorov equation, a partial differential equation of reaction-diffusion type for the tumor cell density $c(\bm{r},t)$ as a function of location $\bm{r}$ and time $t$:
\begin{equation}
\frac{\partial}{\partial t}c(\bm{r},t)=\nabla \cdot \left( D(\bm{r}) \nabla c(\bm{r},t) \right) + \rho c(\bm{r},t) \left( 1-\frac{c(\bm{r},t)}{c^{max}} \right)
\label{EqFKE}
\end{equation} 
where $\rho$ is the proliferation rate which is assumed to be spatially constant, and $D(\bm{r})$ is the $3\times 3$ diffusion tensor which depends on location $\bm{r}$. The first term on the right hand side of equation (\ref{EqFKE}) is the diffusion term that models tumor cell migration into neighboring tissue. The second term is a logistic growth term that describes tumor cell proliferation. The tumor cell density $c(t,\bm{r})$ takes values between zero and the carrying capacity $c^{max}$. In this paper, the diffusion tensor is constructed as
\begin{equation}
D(\bm{r}) = \left\{ \begin{array}{cc}
D_w \cdot I  & \bm{r} \in \mbox{white matter} \\
D_g \cdot I & \bm{r} \in \mbox{gray matter} 
\end{array} \right.
\label{EqD}
\end{equation}
where $I$ is the $3\times 3$ identity matrix, and $D_g$ and $D_w$ are scaling coefficients for gray and white matter, respectively. This construction of the diffusion tensor allows for the modeling of reduced gray matter infiltration via a different diffusion coefficient $D_g < D_w$. Anatomical constraints are accounted for by boundary conditions: We assume that tumor cells only spread in brain tissue and do not diffuse into any tissue that is not segmented as gray or white matter. 
\footnote{The model can be extended towards anisotropic tumor cell migration within white matter. This is achieved by replacing the identity matrix $I$ in equation (\ref{EqD}) by a tensor that is constructed based on the water diffusion tensor obtained through DTI imaging. Although this idea has been conceptually introduced \cite{Clatz2005,jbabdi05}, it is related to several difficulties that are outside the scope of this paper. Problems related to DTI in this context include: (1) even though histopathology provides anecdotal evidence for preferential spread along white matter fibers, there is no data that quantitatively evaluates this effect; (2) the DTI signal is corrupted within the edematous region where the increased water content leads to reduced fractional anisotropy (FA) and increased apparent diffusion coefficients (ADC).
In this paper, we only consider the effects of anatomical boundaries and reduced infiltration of gray matter, because these effects are widely agreed upon, whereas the use of DTI is more speculative at this stage.}

\subsection{Qualitative features of the solution}
The asymptotic solution of the Fisher-Kolmogorov equation is given by a traveling wave solution. At the time of diagnosis, the tumor cell density takes high values close to the carrying capacity $c^{max}$ within the enhancing tumor mass seen on the post-gadolinium T1 image. This is motivated by the observation that resected tissue from this area appears as only tumor tissue in a pathological examination \cite{coons99}. At large distance from the enhancing core, the tumor cell density approaches zero. It can be shown that, within the traveling wave approximation, the following two properties hold:
\begin{itemize}
\item[1.] The fall-off of the cell density with distance from the enhancing core is described by a characteristic profile that remains approximately fixed while the tumor grows \cite{murray02,konukoglu10}. At some distance, the tumor cell density drops approximately exponentially with distance from the visible tumor mass. For a region comprising white matter, the cell density behaves according to
\begin{equation}
c(\vert \bm{r} \vert) \propto \exp \left( -\frac{\vert \bm{r} \vert}{\lambda_w} \right)
\end{equation}
where $\vert \bm{r} \vert$ denotes distance from the visible tumor. The parameter $\lambda_w$ is called the infiltration length and denotes the distance at which the cell density drops by a factor of $e$. It can be shown that $\lambda_w$ relates to the model parameters via
\begin{equation}
\lambda_w=\sqrt{\frac{D_w}{\rho}}
\label{EqLength}
\end{equation}
Within gray matter the cell density drops with an infiltration length $\lambda_g=\sqrt{\frac{D_g}{\rho}}$ which is believed to be smaller than the infiltration length in white matter. 
\item[2.] Over time, the tumor front moves outwards into the brain tissue at a constant velocity. The velocity relates to the model parameters via
\begin{equation}
v_w=2\sqrt{D_w \rho}
\label{EqVelocity}
\end{equation}
\end{itemize}
In a realistic brain geometry, the propagation of the wave is spatially modulated by anatomical boundaries and by different velocities in gray and white matter.

\subsection{Relating the tumor model to imaging data}
In order to apply the model to an individual patient, we have to specify how the tumor cell density resulting from the model relates to the MR images. The contrast enhancing region visible in the T1 post gadolinium MR image is related to a breakdown of the blood-brain-barrier (BBB). Even though this only represents a surrogate for the tumor, it is commonly assumed that contrast enhancing volume corresponds to the vascularized, highly cellular part of the gross tumor. For this paper, we therefore assume that the boundary of the contrast enhancing region corresponds to an isoline of the tumor cell density close to the carrying capacity. It is 
used as the basis for tumor density calculations is described in subsection \ref{SecEnderPaper}. 

In current clinical practice, radiotherapy planning for GBM is mostly based on structural MRI imaging (T1 post gadolinium and T2-FLAIR). However, modern imaging modalities including MR spectroscopy (MRS) and  positron emission tomography (PET) have been used for glioblastoma imaging and bear potential to improve the characterization of gross and infiltrative disease. In particular, tracers visualizing amino acid metabolism have been studied \cite{grosu2005methyl,miwa2004discrepancy,niyazi2011fet,goetz13}, including FET (Fluoro-Ethyl-Tyrosine) and MET (Methionine).  These PET scans are not routinely acquired for glioblastoma patients at our institution. However, if these modalities become established, the tumor growth model could be initialized using a PET based delineation of the tumor instead of using the contrast enhancing region.



\subsection{Calculation of the tumor cell density}
\label{SecEnderPaper}
For treatment planning, we want to calculate the tumor cell density in the brain at the time the MR images are taken. Using the Fisher-Kolmogorov equation directly is problematic in that context because the initial condition, which leads to a tumor consistent with the images, is unknown. One approach to circumvent this problem consists of using the traveling wave approximation. It is assumed that, at the time of imaging, the tumor cell density is characterized by its asymptotic solution. In this approach, it is assumed that the tumor cell density at the T1 gadolinium outline corresponds to an isoline of the tumor cell density. For this paper, we assume a value of 70\% of the carrying capacity. Starting at the outline, the tumor cell density is extrapolated into the brain tissue, assuming the fall-off profile from the traveling wave approximation, and taking into account the anatomical constraints from the brain tissue segmentation. Mathematically, this is achieved by approximating the partial differential equation in (\ref{EqFKE}) by a static Hamilton-Jacobi equation that is solved using a Fast-Marching method. The details of this approach are described in \cite{konukoglu10}.

\begin{figure}[hbt]
\centering
\subfigure[]{
\includegraphics[height=6cm]{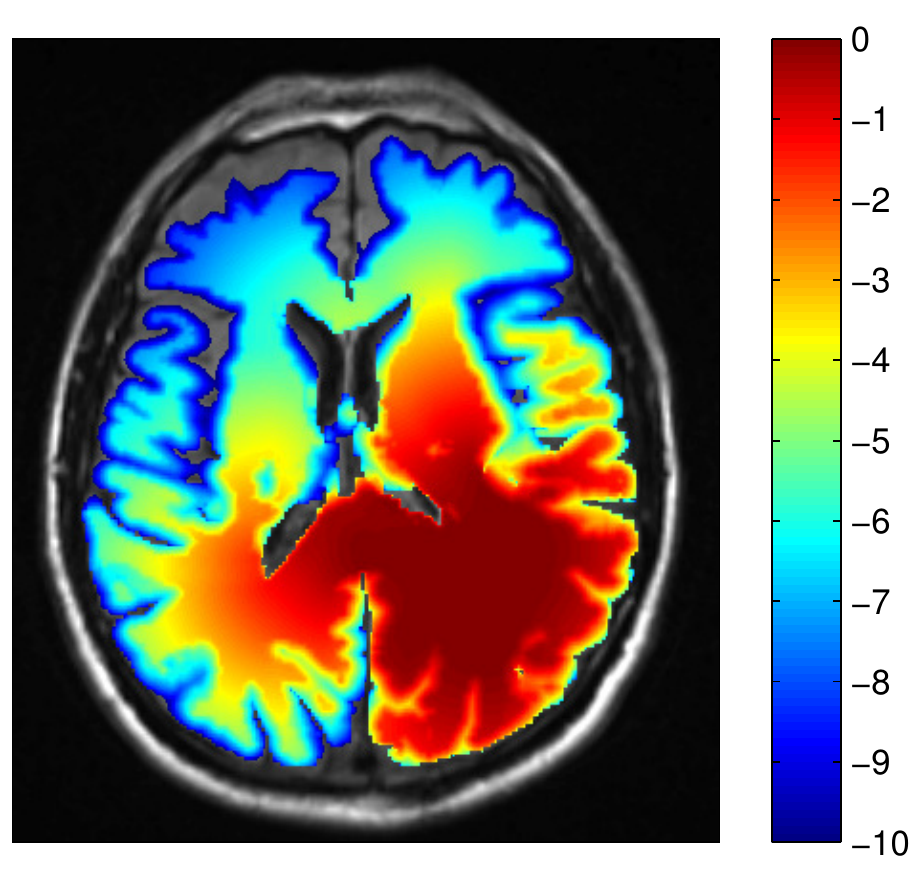}
}
\subfigure[]{
\includegraphics[height=6cm]{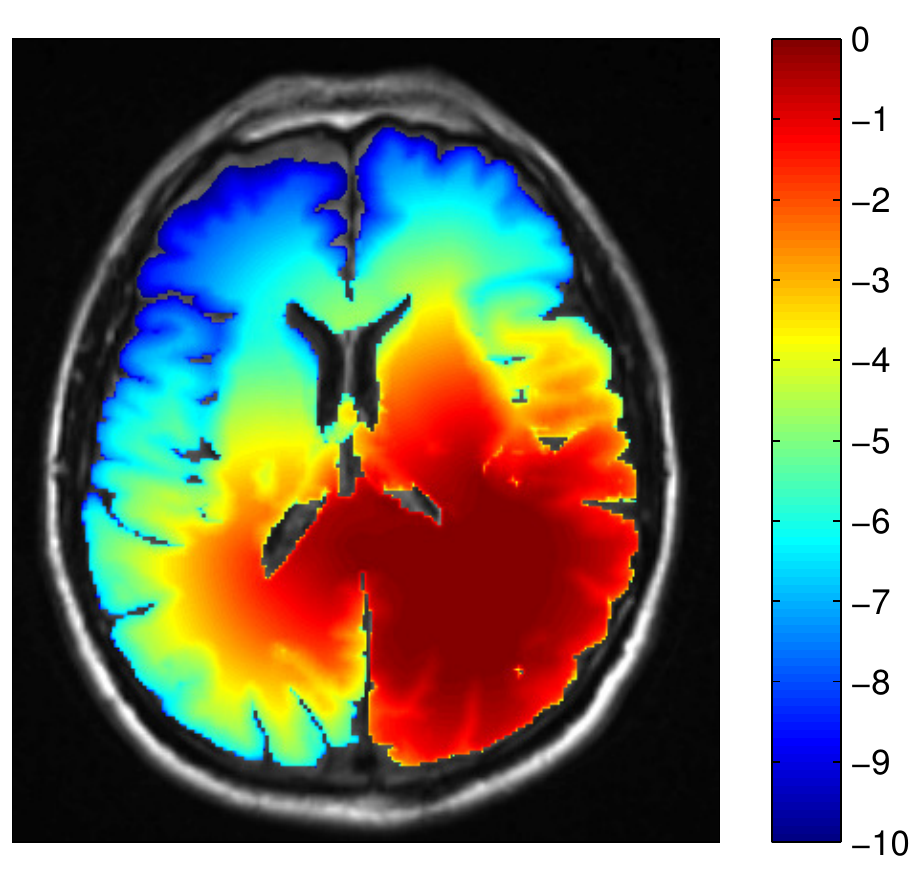}
}
\caption{Simulated tumor cell density based on the segmentation in figure \ref{FigSeg} for parameters $\lambda_w = 4.2$ mm and $D_w/D_g = 100$ (a) and $D_w/D_g = 10$ (b) . The color scale refers to the $log_{10}$ cell density normalized to the carrying capacity.}
\label{FigLogCells}
\end{figure}

\subsection{Model parameters}
The tumor growth model depends on four parameters: The diffusion coefficients in white matter $D_w$ and gray matter $D_g$, the proliferation rate $\rho$, and the carrying capacity $c^{max}$. We now argue that, for our application in radiotherapy planning, the number of parameters can be reduced to two crucial parameters. 

The carrying capacity $c^{max}$ can be considered as a scaling factor for the relative normalized tumor cell density that takes values between zero and one. It is straightforward to see that all results presented in this paper are independent of the value of $c^{max}$ and only depend on the relative tumor cell density. We therefore do not have to consider the value of the carrying capacity $c^{max}$.

We are left with the model parameters $D_w$, $D_g$, and $\rho$. Within the traveling wave approximation, $D_w$ and $\rho$ can be expressed through 1) the velocity of the tumor front in white matter obtained from the product of $D_w$ and $\rho$ according to equation (\ref{EqVelocity}), and 2) the infiltration length in white matter obtained from the ratio of $D_w$ and $\rho$ according to equation (\ref{EqLength}). In this paper, we are only interested in obtaining the distribution of tumor cells at the time when treatment starts. Thus, we do not require the velocity of the tumor front but only the infiltration length $\lambda_w = \sqrt{D_w/ \rho}$. Since the tumor primarily grows in white matter, we focus on the infiltration length in white matter instead of gray matter. If we express the diffusion coefficient in gray matter via the ratio of white and gray matter coefficient, we are left with two parameters of the tumor growth model:
\begin{itemize}
\item[] $\lambda_w = \sqrt{\frac{D_w}{\rho}}$ : Infiltration length, describing how fast the tumor cell density drops with distance from the visible tumor volume.
\item[] $D_w / D_g$ : Ratio of diffusion coefficients in white matter and gray matter, parameterizing in parts the anisotropy of tumor growth. The ratio of the infiltration lengths is consequently given by $\lambda_w /\lambda_g = \sqrt{D_w / D_g}$.
\end{itemize}

In this paper, we are concerned with the spatial definition of the target volume as an isoline of the tumor cell density. In that context, the exact value of the infiltration length $\lambda_w$ is irrelevant because it does not affect the shape of the isolines. Only the abolute values of the tumor cell density associated with the isolines are determined by $\lambda_w$ \footnote{The value of $\lambda_w$ will be crucial for dose prescription considered in the accompanying paper \cite{unkelbach13b}.}. Thus, the only relevant model parameter for this paper is the ratio $D_w/D_g$, which determines the shape of the isolines of the cell density together with the brain segmentation. The literature consistently suggests that tumor cells infiltrate gray matter much less than white matter\footnote{This holds for the most common case of astrocytomas, not necessarily for oligodentrogliomas (see \cite{coons99} for a review of glioma growth patterns).} This suggests a large value for $D_w/D_g \gg 1$. Most illustrations in this paper were obtained for $D_w/D_g = 100$. The most appropriate value is however uncertain and we discuss the impact of uncertainties in $D_w/D_g$ in section \ref{SecSensitivity}.

\section{Model based target delineation}
\label{SecCase}

In this section we demonstrate the use of the tumor growth model for target definition. 
In sections \ref{SecReshape}-\ref{SecSensitivity}, we provide a detailed discussion of the patient shown in figure \ref{FigMRI} in order to illustrate the main findings of the study. We first compare the model-derived target volumes to the manual delineation used in the clinical treatment plan. We then discuss the impact of anatomical boundaries and reduced gray matter infiltration in more detail, and finally discuss sensitivity to model inputs. We analyzed 10 GBM patients with varying tumor locations previously treated at our institution. In section \ref{SecOtherCases}, we summarize the main findings for these patients; further details are provided in \ref{SecAppendix}.

\subsection{Target volume definition}
\label{SecReshape}
\begin{figure}[hbt]
\centering
\includegraphics[height=6cm]{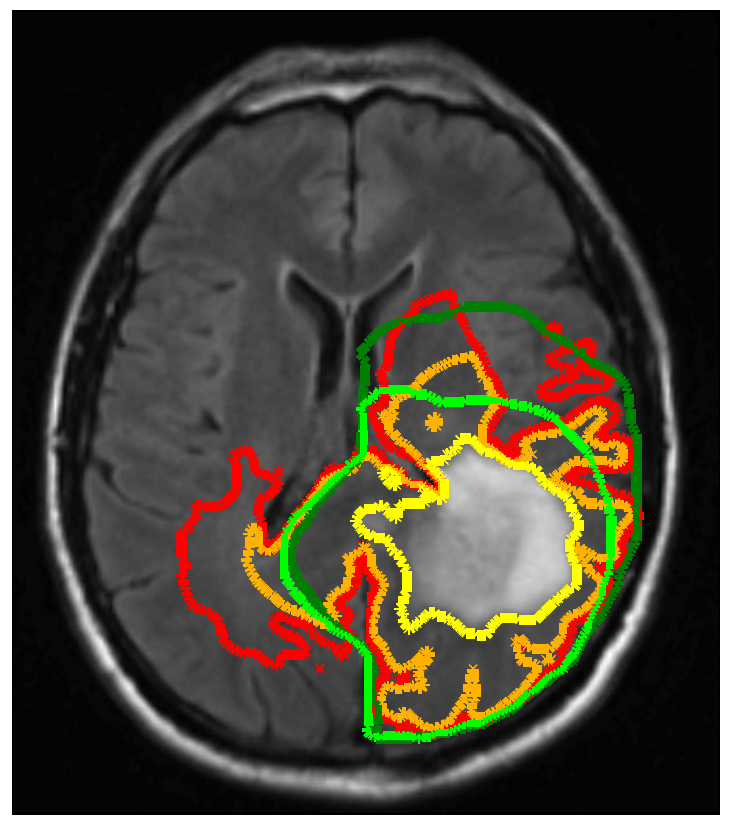}
\caption{Comparison of manually defined targets (light/dark green) and model derived targets (orange/red). In the clinical treatment plan, the light green volume was prescribed to 60 Gy, the dark green volume to 46 Gy. The yellow contour shows the abnormality on T2-FLAIR.}
\label{FigTarget}
\end{figure}

Figure \ref{FigTarget} shows the clinical target volume (dark green contour) and the boost volume (green contour) drawn manually by the physician, i.e. these contours were the basis for the treatment plan that was actually delivered to the patient. The prescribed dose to the boost volume was 60 Gy, the prescribed dose to the CTV was 46 Gy. The boost volume is defined based on a 2 cm isotropic extension of the contrast enhancing lesion. The CTV is defined via a 1.5 cm expansion of the T2-FLAIR abnormality. Both volumes were subsequently trimmed manually to account for anatomical boundaries (dura, ventricles, falx, and tentorium cerebelli). The dose distribution of the 3D conformal treatment plan delivered to the patient is shown in figure \ref{FigDose}d.\\

Using the tumor growth model, the target volume can be defined as an isoline of the tumor cell density. This is illustrated here for the tumor cell density shown in figure \ref{FigLogCells}a obtained for the parameter value $D_w/D_g=100$. The red and the orange contours in figure \ref{FigTarget} show the CTV and the boost volume derived from the tumor growth model, respectively. In this example, the target defining isolines are chosen such that the total enclosed volume is equal to the manually delineated target. In the following two subsections, we discuss the differences between manual and model-derived target volumes in detail.

\subsection{Accounting for anatomical boundaries}
\label{SecBarriers}
\begin{figure}[hbt]
\centering
\subfigure[]{
\includegraphics[height=6cm]{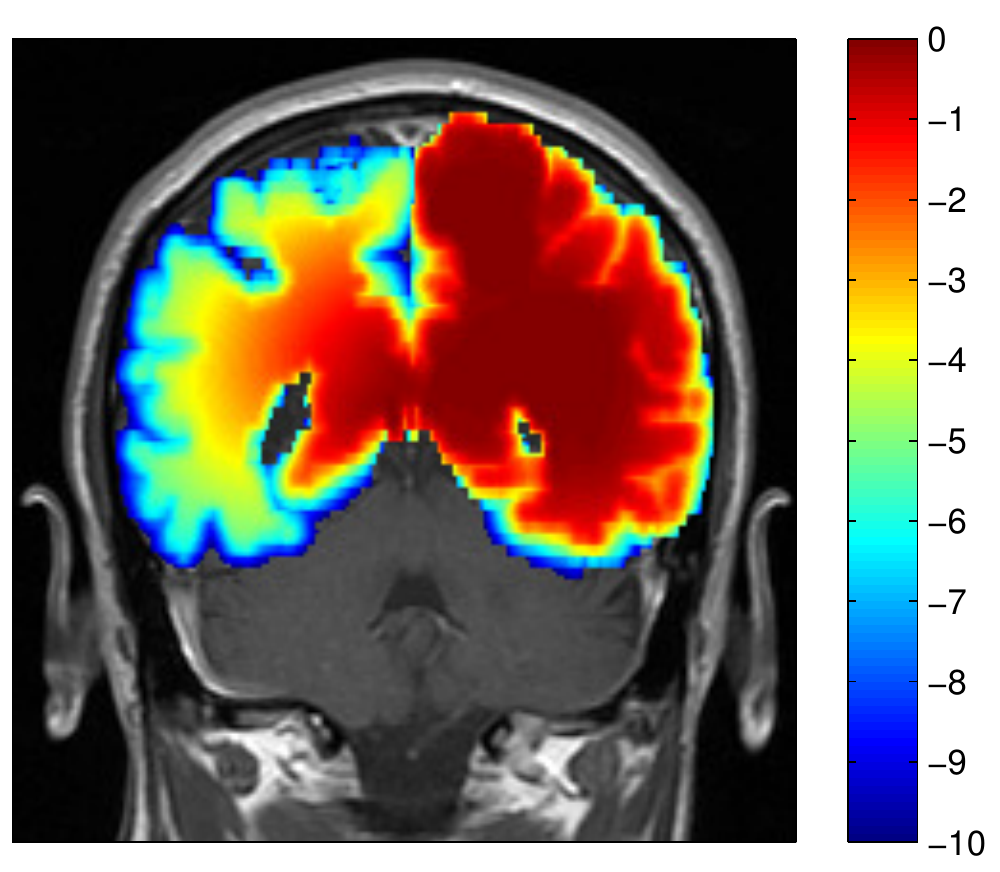}
}
\subfigure[]{
\includegraphics[height=5.8cm]{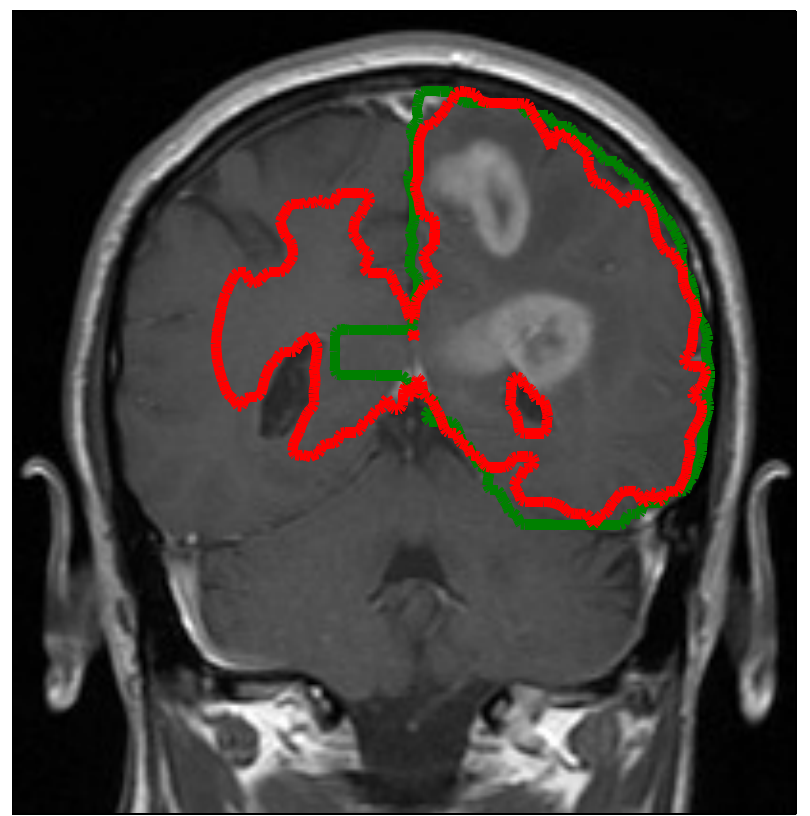}
}
\caption{(a) Log tumor cell density overlayed on the coronal T1 image showing spread of the tumor in the contralateral hemisphere ($\lambda_w = 4.2$ mm and $D_w/D_g = 100$). (b) Comparison of manual CTV (green) and model based CTV (red) (same contours as in figure \ref{FigTarget}.}
\label{FigCoronalCells}
\end{figure}

In the manual delineation of the CTV used in the clinical plan, it is incorporated that the falx represents an anatomical barrier for the migration of tumor cells. Hence, the isotropic target expansion was trimmed manually. In the tumor growth model, the falx is modeled via a layer of CSF and is automatically accounted for through the assumption that tumor cells only migrate within white and gray matter. However, the corpus callosum connects the two hemispheres of the brain. The tumor growth model describes the migration of tumor cells through the corpus callosum (see figure \ref{FigLogCells} and \ref{FigCoronalCells}a). As a consequence, the target volume based on the growth model is extended into the contralateral hemisphere. Figure \ref{FigCoronalCells}a shows the tumor cell density overlaid on the coronal T1 gadolinium image. This illustrates the three-dimensional modeling of tumor spread via the model, including areas superior to the corpus callosum. This is not consistently accounted for in the manual CTV. In the manually drawn target volumes, the target is slightly extended into the contralateral hemisphere on the slices that show the corpus callosum, but not on the slices located superiorly and inferiorly (figure \ref{FigCoronalCells}b). In the model derived target volumes, the target is extended further into the contralateral hemisphere, and the spread of tumor cells in superior-inferior direction beyond the corpus callosum is modeled.

\subsection{Modeling reduced infiltration of gray matter}
\label{SecGrayMatter}
It is typically assumed that tumor cells infiltrate gray matter much less than white matter, which is modeled by $D_w/D_g \gg 1$. In the simulated tumor cell density in figure \ref{FigLogCells}, this leads to a sharp drop of the cell density towards the cortex. In most regions, this effect does not alter the global shape of the target volume substantially because the thickness of the cortical layer is only several millimeters. This can be seen in figure \ref{FigLogCells} in the posterior portion of the tumor. Here, the dura is essentially the boundary of the target volume. Technically, some gray matter surrounding the fissures is outside of the target defining isoline. However, these regions are small and are not relevant for radiotherapy planning as further discussed in section \ref{SecPlanning}. For this patient, the most pronounced differences between the model derived target and the manually drawn target, which can be attributed to reduced gray matter infiltration, arise in the region of the lateral sulcus (left-anterior part of the tumor). In this region, the large amount of gray matter surounding the lateral sulcus represents a soft barrier for migrating tumor cells. Thus, the tumor growth model can be used to identify regions of functioning brain tissue that can be taken out of the target volume.

\subsection{Difference in target volumes}
\label{SecDiceTarget}
In order to quantitatively compare the difference of manual and model-derived target volumes, we calculate the Dice coefficient, which is given by the volume where both structures overlap, divided by the average volume of the two structures. For the boost and CTV volumes in figure \ref{FigTarget} the Dice coefficients are 0.78 and 0.77, respectively. Thus, three quarters of the manual target is also contained in the model-derived target, but a substantial portion of approximately one quarter is different.

\subsection{Sensitivity to input parameters}
\label{SecSensitivity}
The model input parameters that determine the spatial distribution of tumor cells are the segmentation of the brain (into white matter, gray matter and CSF) as well as the parameter $D_w/D_g$. 

\paragraph{Brain segmentation:} Anatomical boundaries like the ventricles and the falx are modeled via boundary conditions. As a consequence, correct segmentation of the brain is a crucial input to the model. This can be illustrated for the falx: Correct modeling of the anatomical boundary in our approach relies on identifying a layer of CSF along the falx. If due to limitations in the automatic segmentation, the two hemispheres are connected through white or gray matter bridges (outside of the corpus callosum), the falx is not correctly established as a barrier. In this case, the growth model will incorrectly predict interhemispherical cell migration outside of the corpus callosum. This problem can in particular arise if, due to a tumor with mass effect, the brain tissue is compressed and the falx is pushed towards the contralateral hemisphere. A reliable application of the model for target definition therefore requires reliable segmentation of the ventricles and the extensions of the dura, i.e. falx cerebri and tentorium cerebelli. In addition, correct modeling of contralateral spread requires a reliable segmentation of the corpus callosum as white matter. This can, for example, be problematic in the superior region of the corpus callosum when the slice distance in the MRI scan is large. Approaches to customize brain segmentation methods for this application can be found in \cite{unkelbach12b}.

\paragraph{Gray matter infiltration parameter:} In addition to the segmentation, the ratio of white and gray matter diffusion coefficient $D_w/D_g$ influences the shape of the isolines of the tumor cell density. For $D_g = 0$, gray matter represents a hard boundary and tumor cells only spread in white matter. For $D_w/D_g = 1$, tumor cells spread equally in white and gray matter and the shape of the target is solely influenced by anatomical constraints. In figure \ref{FigLogCells} the simulated tumor cell density is compared for $D_w/D_g=10$ and $D_w/D_g=100$. For smaller $D_w/D_g$, the cell density is more washed out (figure \ref{FigLogCells}b) compared to a larger $D_w/D_g$ where the tumor cell density follows more closely the white matter structure (figure \ref{FigLogCells}a). It has been discussed above that the cortical gray matter has a thickness of only a few millimeters. As a consequence,  varying $D_w/D_g$ has little impact on the global shape of the target volume. The most significant changes for this patient are around the lateral sulcus.

\paragraph{Interdependence:} There is some interdependence regarding the sensitivity against $D_w/D_g$ versus segmentation errors. For small  values of $D_w/D_g$, the result may be sensitive to errors in the segmentation of CSF, whereas errors in the segmentation between white and gray matter become irrelevant. For large values of $D_w/D_g$, the segmentation of CSF becomes less crucial because falx and tentorium cerebelli are surrounded by a layer of gray matter which models the boundary. However, errors in the segmentation of white matter versus gray matter become more crucial.

\subsection{Target volume comparison for different tumor locations}
\label{SecOtherCases}
We retrospectively analysed 10 patients with varying tumor location. All patients were treated with IMRT to a dose of 60 Gy in 30 fractions. For the clinically delivered treatment plan, the CTV was defined using isotropic expansions of 1-2 cm to the T2-FLAIR hyperintensity. Manual editing was performed to trim the CTV expansion where there were clear anatomical barriers to spread such as the falx or dura. A PTV expansion of 3 mm was added. For all patients, the tumor cell density is calculated for a parameter value of $\dwdg = 100$ based on the method described in section \ref{SecModel}. The model-based target volume is defined as the isoline that encompasses the same total volume as the manually delineated CTV used in the clinically applied plan. The main findings are summarized as follows:
\begin{itemize}
\item[$\bullet$] The modeling of anatomical barriers leads to differences in the target volume for tumors located close to corpus callosum and falx cerebri. The use of the tumor growth model suggests a further expansion of the target into the contralateral hemisphere.
\item[$\bullet$] The modeling of reduced gray matter infiltration leads to differences in the target volume for tumors located in proximity to major sulci. This effect was most pronounced in the region of the lateral sulci where the use of the tumor growth model suggests regions that can be excluded from the target volume.
\item[$\bullet$] Dice coefficients between manual and model derived target volumes were calculated for all 10 patients (table \ref{TabDice}). The overlap between manual and model-based target volumes was 79\% on average, ranging from 74\% to 84\%. This indicates that approximately 20\% of the manual target volume is not contained in the model-based volume (and vice versa).
\item[$\bullet$] For all patients IMRT treatment plans were optimized using a homogeneous 60 Gy prescription dose to the CTV and the planning formulation in section \ref{SecPlanning}. Dice coefficients for the 95\% isodose lines (57 Gy) were calculated in order to compare the differences in the high dose region between manual and model derived radiation plans (table \ref{TabDice}). The Dice coefficient was 0.84 on average,  ranging from 0.80 to 0.87, indicating that a substantial part of the differences in target volume also translates into differences in the high dose region. 
\end{itemize}
In the appendix \ref{SecAppendix}, we provide details for four representative cases. These four cases span a range of different tumor locations, i.e. GBMs in the parietal lobe, temporal lobe, frontal lobe, and bilateral corpus callosum. 

\begin{table}[hbt]
\begin{centering}
\begin{tabular}{l|l|c|c}
& Location & Dice CTV contour & Dice 57 Gy isoline \\
\hline
1 & Parietal (Fig. \ref{FigTarget}) & 0.77 & 0.82 \\ 
2 & Temporal/Frontal  (Fig. \ref{FigPat003}) & 0.74 & 0.81\\ 
3 & Parietal  (Fig. \ref{FigPat004}) & 0.81 & 0.82\\ 
4 & Temporal  (Fig. \ref{FigPat006}) & 0.84 & 0.87\\ 
5 & Corpus Callosum  (Fig. \ref{FigPat015}) &  0.76 & 0.82 \\ 
6 & Temporal  (not shown) &  0.81 & 0.85 \\ 
7 & Temporal (not shown) &  0.83 & 0.86 \\ 
8 & Parietal (not shown) &  0.80 & 0.84 \\ 
9 & Parietal (not shown) &  0.84 & 0.85 \\ 
10 & Temporal (not shown) &  0.74 & 0.80 
\end{tabular}
\caption{Dice coefficients for the CTV contours and the 57 Gy isoline of an IMRT plan for different tumor locations.}
\label{TabDice}
\end{centering}
\end{table}

\subsection{Model-based target delineation without a reference volume}

In figure \ref{FigTarget} we have defined the model derived target contours such that the overall volumes of the targets equal the volumes of the targets in the clinical plan. This has been done for the sake of comparison of the shapes of the target volumes. In a practical scenario, the model is used for target definition without the existence of a manually drawn reference volume. In this case, the choice of the target defining isoline can be performed by either specifying the target expansion in white matter, or by specifying the average expansion:
\begin{itemize}
\item[1.] In the first scenario, the treatment planner is provided with the simulated tumor cell density together with the delineation of enhancing core and edema as well as the brain segmentation. The treatment planner can then manually pick a target defining isoline by measuring the distance to the enhancing core along major sheets of white matter.
\item[2.] In the second case, a temporary reference volume can be created automatically as an isotropic expansion around the enhancing core, intersected with the brain tissue mask. The model-derived target volume can then be defined as the isoline that encompasses the same total volume as the temporary reference volume. In this case, the treatment planner has to specify the average margin added to the enhancing core.
\end{itemize}

\section{Treatment plan comparison}
\label{SecPlanning}
In this section we illustrate the impact of the model derived target volumes on the dose distribution. A treatment plan for intensity modulated radiotherapy was optimized, using 9 equally spaced coplanar 6MV photon beams. Dose calculation was performed using the Quadrant Infinite Beam (QIB) algorithm in CERR \cite{deasy2003cerr}, and the optimization of IMRT treatment plans was performed using our own implementation of the L-BFGS quasi-newton method. More specifically, we minimize the following piece-wise quadratic objective function
\begin{eqnarray}
f(d) &=& \sum_\eta \frac{w^o_\eta}{N_\eta}\sum_{i \in V_\eta} \left( d_i -d^{max}_\eta \right)_+^2 \label{EqObjOar} \\
&+& \frac{w^u_T}{N_T} \ \sum_{i \in T} \left( d^{pres}_i - d_i \right)_+^2  
+ \frac{w^o_T}{N_T} \ \sum_{i \in T} \left( d_i - d^{max}_T \right)_+^2 \label{EqObjTarget} \\
&+& \frac{w^o_H}{N_H} \sum_{i \in H} \left( d_i - d^{max}_i \right)_+^2 \label{EqObjFalloff}
\end{eqnarray}
The first term (\ref{EqObjOar}) denotes overdose objectives for the organs at risk (OAR). For the maximum doses $d^{max}_\eta$ and weighting factors $w_\eta^o$, we use the values summarized in table \ref{TabIMRT}. The second term (\ref{EqObjTarget}) represents under and over dose objectives for all target voxels. The prescribed dose $d_i^{pres}$ is set to 60 Gy for all voxels $i$ in the boost volume and 46 Gy in the CTV. The maximum dose $d^{max}_T$ in all target voxels is set to 60 Gy. For all unclassified normal tissues surrounding the target volume (including skull, brain tissue, ventricles), we use a conformity objective in which $d^{max}_i$ decreases linearly with the distance of voxel $i$ from the target volumes. Here, we define a generalization of the conformity objective to the case of an inhomogeneous prescription dose. More specifically, for every voxel $i$ we define 
\begin{equation}
d^{max}_i = \max \left[ d^{low} , \max_j \left( d^{pres}_j - z_{ij} \, d^{grad} \right) \right] \\
\end{equation}
where $z_{ij}$ denotes the euclidean distance of the voxel $i$ from another voxel $j$ that has a non-zero prescription dose $d^{pres}_j$. The parameter $d^{grad}$ is specified in Gy per cm and describes the desired dose falloff in healthy tissue; $d^{low}$ is a lower dose threshold below which dose is not penalized. The optimization parameters used are summarized in table \ref{TabIMRT}.

\begin{table}[hbt]
\begin{centering}
\begin{tabular}{l|c|c|c|c|c|c}
& $w^o$ & $w^u$ & $d^{max}$ & $d^{pres}$ & $d^{grad}$ & $d^{low}$ \\
\hline
CTV & 10 & 20 & 60 & 46 &-&-\\
Boost & 10 & 20 & 60 & 60 &-&-\\
\hline
Brainstem & 10 & - & 45 & - &-&-\\
Optic nerves & 10 & - & 30 & - &-&-\\
Chiasm & 10 & - & 30 & - &-&-\\
Eye lenses & 10 & - & 10 & - &-&-\\
Eye balls & 10 & - & 10 & - &-&-\\
Unclassified & 10 & - & - & - & 40 & 20
\end{tabular}
\caption{Objective function parameters used for IMRT optimization}
\label{TabIMRT}
\end{centering}
\end{table}

\begin{figure}[hbtp]
\centering
\subfigure[]{
\includegraphics[height=6cm]{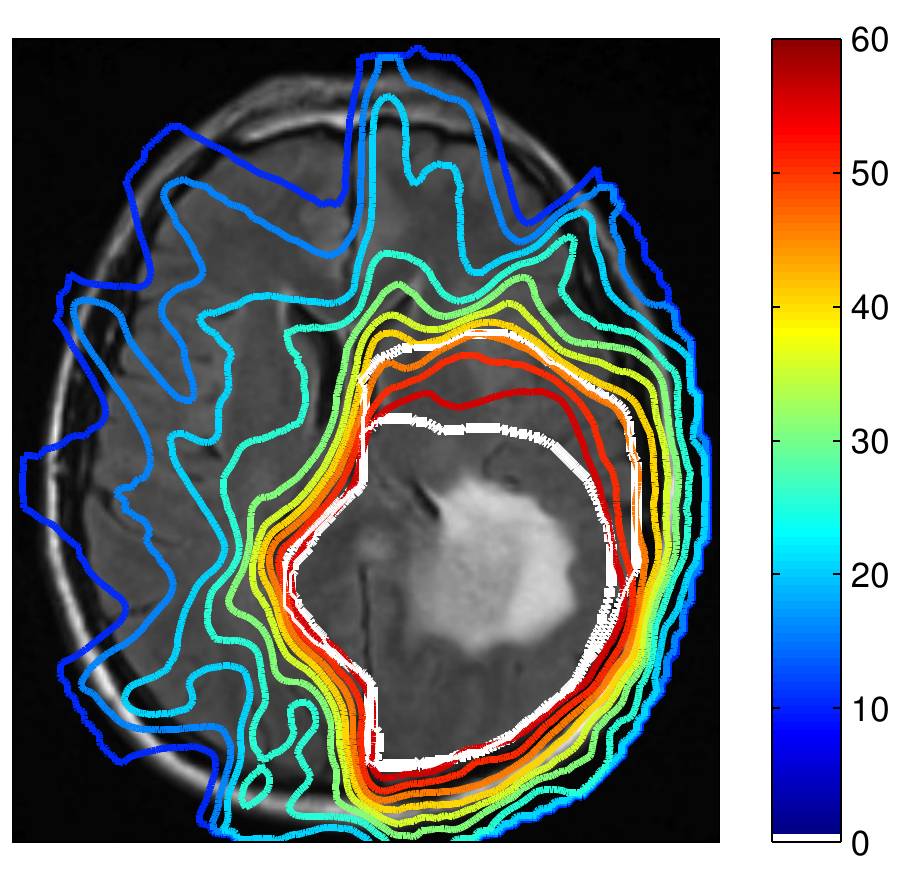}
}
\subfigure[]{
\includegraphics[height=6cm]{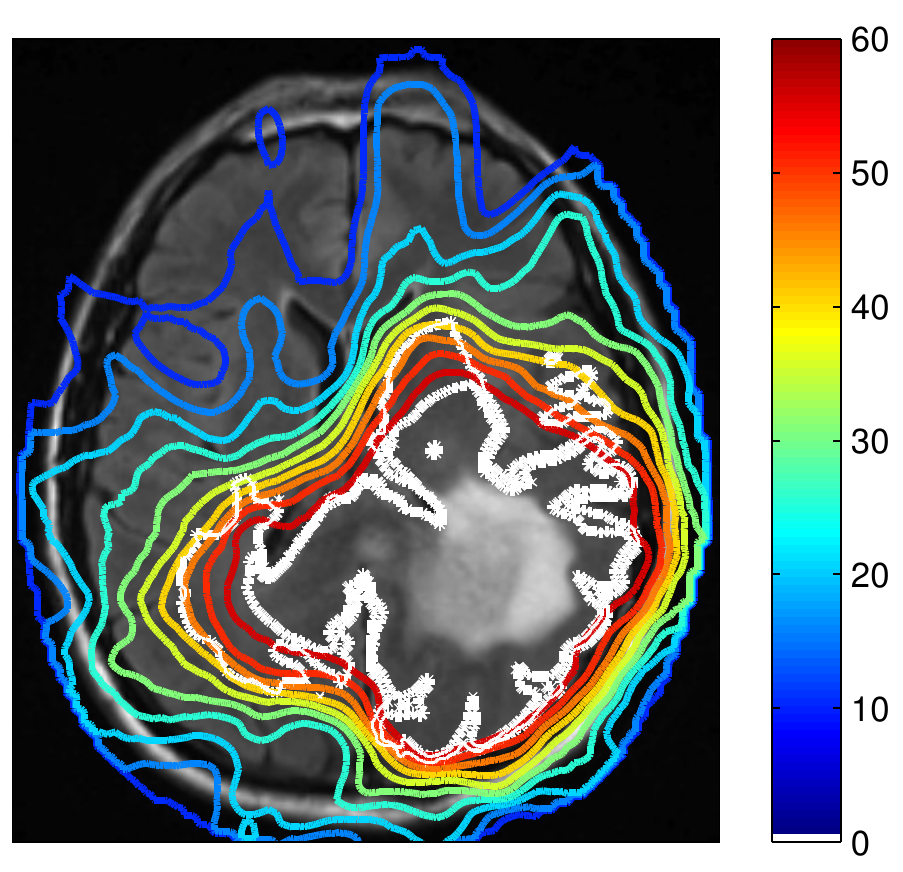}
}
\subfigure[]{
\includegraphics[height=6cm]{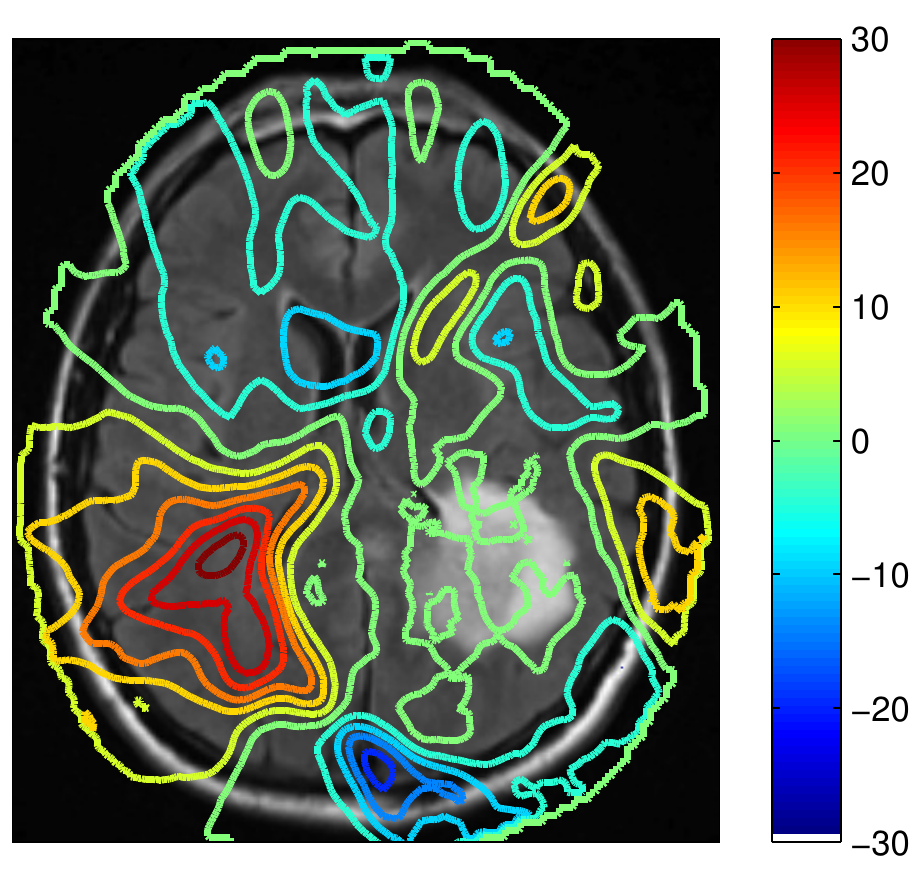}
}
\subfigure[]{
\includegraphics[height=6cm]{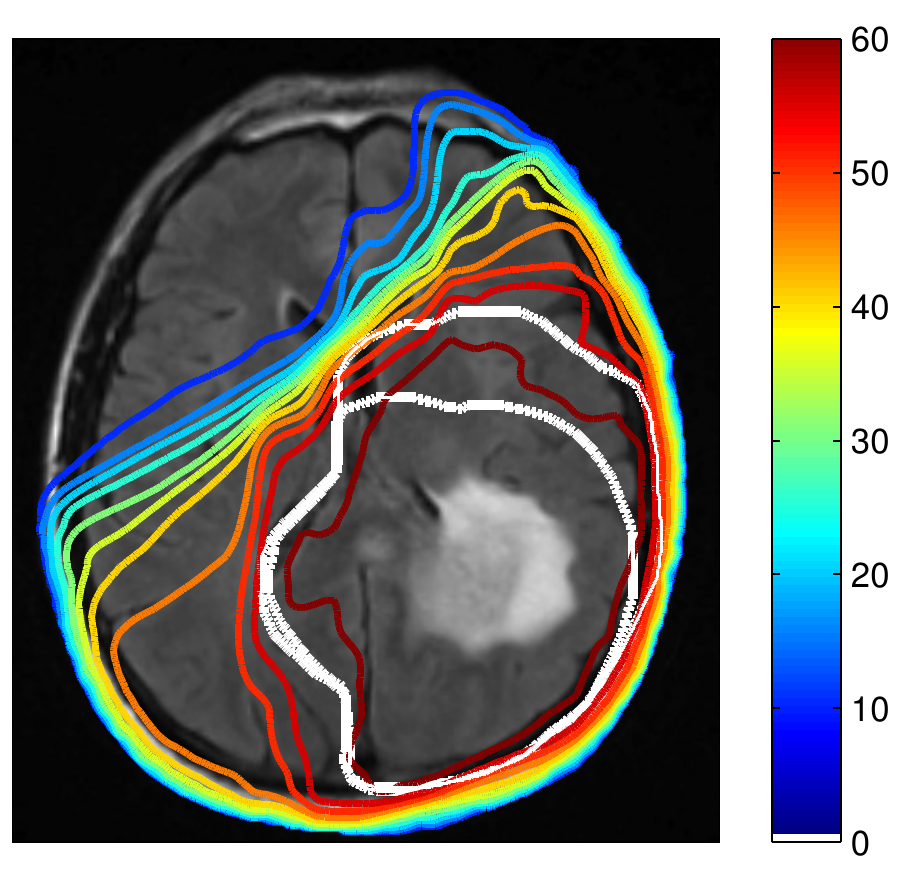}
}
\caption{Comparison of dose distributions (in units of Gy): (a) for a plan based on manual target volumes; (b) for model-derived target volumes. The difference of the dose distribution is shown in (c). Figure (d) shows the planned dose distribution for the 3D conformal plan the patient was treated with.}
\label{FigDose}
\end{figure}

Figure \ref{FigDose} shows the dose distributions obtained for two treatment plans. The plan in figure \ref{FigDose}a is based on the manually drawn target volumes, whereas the plan in figure \ref{FigDose}b is based on the target volumes derived from the growth model. For the sake of comparison, the plan based on model-derived targets was optimized by imposing an additional constraint on the integral dose delivered to the patient, i.e. we minimize objective function (\ref{EqObjOar}-\ref{EqObjFalloff}) subject to the constraint
\begin{equation}
\sum_{i \in V} d_i \leq d^{int}
\end{equation}
where $V$ denotes the set of all voxels in the patient, and $d^{int}$ denotes the integral dose obtained for the treatment plan based on the manual target volumes. Figure \ref{FigDose}c shows the difference of the plans in figures \ref{FigDose}a and \ref{FigDose}b. It can be seen that in plan \ref{FigDose}b more dose is delivered to the contralateral hemisphere. Under the constraint that integral dose is not increased, this is compensated for by delivering less dose to other regions, e.g. near the lateral sulcus. \\

To quantify the difference in dose distributions in figure \ref{FigDose}, we calculate the Dice coefficients for the volumes enclosed by selective isodose lines. The Dice coefficients for the 57 Gy and the 46 Gy isodose line evaluate to 0.84 and 0.83, respectively. In section \ref{SecDiceTarget} we calculated the Dice coefficient between manual and model-derived boost and CTV volumes, which are 0.78 and 0.77, respectively. It is apparent that differences in target volumes only partly translate into differences in dose distributions. The rugged shape of the model-derived target volume that arises from reduced infiltration of gray matter is mostly not of relevance for radiotherapy planning. Due to the physical characteristics of photon beams it is not possible to spare thin layers of gray matter surrounding small fissures.

\section{Comparison to follow-up imaging}
\label{SecValidation}

\subsection{Scope}
Ideally, we would like to have a way to validate the predictions of the growth model regarding the tumor cell density. Histopathological analysis of resected brain tissue would represent one way to validate the extent of tumor cell infiltration predicted by the model. However, it is currently not realistic to obtain such data for untreated human patients at high spatial resolution. Tissue samples taken from biopsies or resected tumors may provide information at selected locations. Usually this is, however, limited to the gross tumor volume; taking biopsy samples from healthy appearing brain tissue, which would be relevant for target delineation, is not performed.

An alternative approach towards collecting evidence for the plausibility of the growth model consists in evaluating the spatial progression patterns of recurrent tumors in follow-up MRI images. The advantage of this approach is that this data is widely available since high grade glioma patients typically receive follow-up MR imaging every 2-3 months. On the downside, model validation based on follow-up imaging has limitations. This includes the difficulty in distinguishing recurrent tumor from other radiation induced changes (e.g. radiation necrosis), and the large variation in response to radiation between patients. In addition, deformations due to surgery, tumor mass effects, and radiation related brain atrophy may occur. Addressing these issues in their complexity is outside the scope of this paper. Nevertheless, we hypothesize that the analysis of progression patterns of recurrent tumors can provide some degree of evidence for the plausibility of the tumor growth model. Due to the variability in treatment and treatment response, this is not the case for all patients - but for individual cases.

\subsection{Patient characteristics}
\begin{figure}[hbt]
\centering
\subfigure[3.5 month]{
\includegraphics[height=6cm]{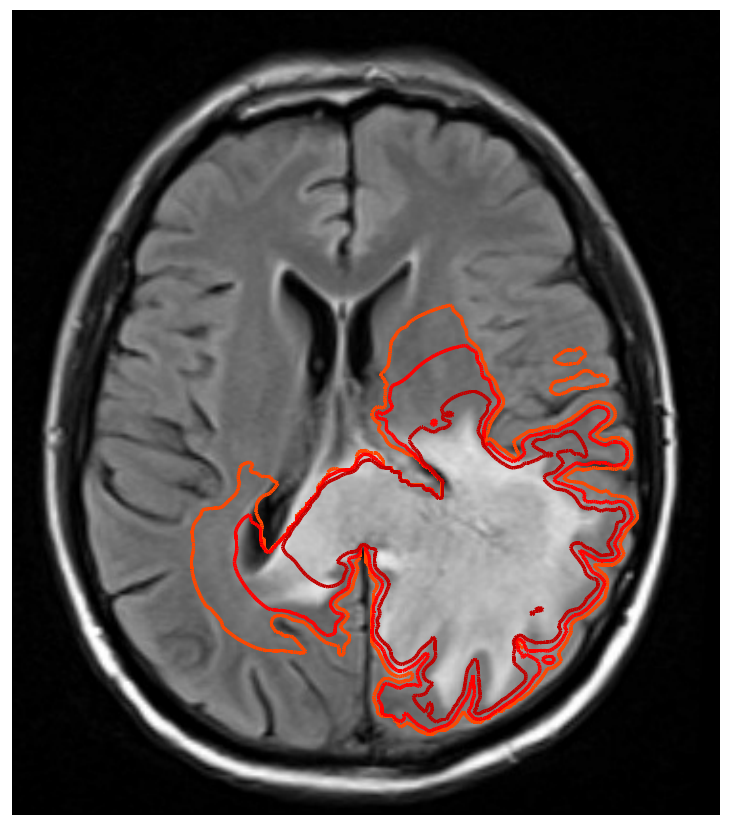}
}
\subfigure[7.5 months]{
\includegraphics[height=6cm]{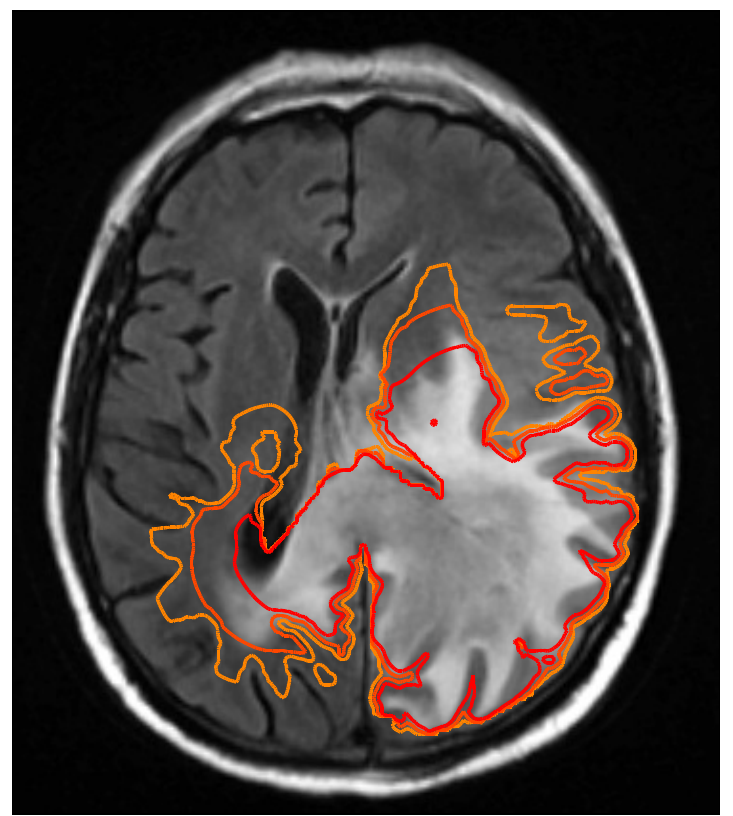}
}
\caption{T2-FLAIR images acquired after therapy, overlaid with isolines of the tumor cell density simulated on the diagnostic images prior to radiotherapy (figure \ref{FigLogCells}a): (a) 3.5 month after diagnosis with isolines for tumor cell densities of $10^{-1}$, $10^{-2}$, and $10^{-3}$, and (b) 7.5 months after diagnosis with isolines for tumor cell densities of $10^{-2}$, $10^{-3}$, and $10^{-4}$.}
\label{FigFollowup}
\end{figure}

The patient discussed in sections \ref{SecCase} and \ref{SecPlanning} only had a biopsy taken prior to radiotherapy without significant tumor resection. Also, the patient did not receive additional cycles of chemotherapy or antiangiogenic drugs after the initial course of chemoradiation with concurrent temozolomide. Figure \ref{FigFollowup} shows two follow-up T2-FLAIR images for this patient. The follow-up images were registered to the reference image using rigid registration. The image in figure \ref{FigFollowup}a was taken 3.5 months after the diagnostic image (figure \ref{FigMRI}) and 2 months after completion of radiotherapy. The image in figure \ref{FigFollowup}b was taken 7.5 months after the diagnostic image, which was the last image taken before the patient expired. 
Even though there is no histopathological confirmation, the radiological appearance and the patient's death about 9 months post therapy suggest progressive disease rather than other radiation induced changes. Thus, for the remainder of this section, we make the assumption that the follow-up imaging shows the progression of the tumor. Furthermore, the tumor recurs/progresses centrally, which is the case for the vast majority of cases given standard of care treatments.
Figure \ref{FigDose}d shows the dose distribution of the clinically delivered 3D conformal treatment plan. 

\subsection{Comparison of follow-up images and simulated tumor infiltration}

By considering the delivered dose distribution in figure \ref{FigDose}d, we see that the progressive tumor is located almost entirely within the high dose region receiving approximately 60 Gy. Under the assumption that a radiation dose of 60 Gy kills a fraction of the tumors cells, the shape of the isolines after therapy is not altered within the high dose region. In this case, the Fisher-Kolmogorov model predicts that the progression of the tumor approximately follows the isolines of the tumor cell density simulated pre therapy. This motivates to compare the simulated cell density isolines directly to the follow-up images without explicitly modeling the effect of chemoradiation on the cell density quantitatively.

In figure \ref{FigFollowup} we overlaid the tumor cell density simulated on the diagnostic image on the follow-up T2-FLAIR images. It is seen that the progression of the tumor post therapy follows the isolines of the simulated tumor cell density relatively closely. In particular, the follow-up images show the progression of the tumor through the corpus callosum into the contralateral hemisphere as predicted by the model. The latter is supported by the T1 post contrast images 3.5 and 7.5 months post therapy, which show contrast enhancement in the corpus callosum (T1 images not shown).

\section{Discussion}
\label{SecDiscussion}

\subsection{The problem addressed in this work}
In this work, we aim at improving and automating target volume delineation for glioblastoma. We investigate the use of a phenomenological tumor growth model based on a reaction-diffusion equation for the tumor cell density. The model has been studied in the literature over the past years. Here, we aim at bringing this model closer to an application in radiotherapy treatment planning. The problem addressed here originates from the infiltrative growth characteristics of glioma and the fact that tumor cells infiltrate brain tissue beyond the abnormality visible on current imaging modalities. In clinical practice, an approximately 2 centimeter wide margin is added around the T2-FLAIR abnormality. It is however known that glioma growth is not isotropic. Tumor growth patterns are influenced by anatomical boundaries as well as reduced infiltration of gray matter. Hence, adding an isotropic margin will unnecessarily expose normal tissue or miss areas of concern. Anatomical constraints can partly be accounted for in the manual definition of the CTV, e.g. by excluding ventricles and regions outside of the dura. However, a consistent modeling of more complex 3D anatomical features is difficult to achieve manually. 

\subsection{Main findings}
We performed a retrospective study involving 10 GBM patients treated previously at our institution. For these patients, we compare manually drawn target volumes used in the clinically delivered treatment plan to model-derived target volumes defined through an isoline of the tumor cell density. One goal of the study is to characterize the best use cases for model based target delineation, i.e. we identify anatomical situations in which the growth model leads to more consistent target volumes compared to manual delineation. As demonstrated in the paper, this may be the case for tumors located in proximity to falx and corpus callosum. In this situation, the falx represents a boundary for migrating cells, but the corpus callosum provides a route for the tumor to spread to the contralateral hemisphere. The tumor growth model represents a tool to automate and objectify target definition for such cases, which is difficult to achieve manually. In addition, it is demonstrated that modeling of reduced infiltration of gray matter may influence target delineation near major sulci, where large accumulations of gray matter surrounding the sulci represent a soft boundary for migrating tumor cells. In particular, this effect is observed for tumors located in proximity to the lateral sulcus. By optimizing IMRT treatment plans based on model-derived and manually drawn target volumes, it is demonstrated that the differences in target delineation may translate into differences in the delivered dose distribution. 


\subsection{Scope of the work} 

\paragraph{Technical versus clinical goal:} The work presented has both a technical goal and a clinical goal. The technical goal of this project is to develop automated delineation of the CTV while accounting for anisotropic growth patters. Currently, anatomical barriers are to some degree taken into account manually. In complex situations, this is a time consuming process and often leads to inconsistent CTVs. Automated GTV to CTV expansion will greatly accelerate treatment planning and make treatment volumes more objective. In addition to the technical goal, more accurate identification of the highest risk regions of microscopic disease may translate into a clinical advantage. The large isotropic CTV expansions currently used include neighboring low risk normal brain tissue, which may receive unnecessary and potentially harmful irradiation. However, by using smaller CTV margins, clinicians risk missing areas of tumor extension, which may lead to earlier locoregional failure. Due to the heterogeneity among GBM patients,  the clinical advantage will be difficult to demonstrate, however, the technical goal of automating CTV delineation is itself worthwhile to pursue. 

\paragraph{The role of model validation:} Common criticisms to the use of tumor growth models for radiotherapy planning are that models are oversimplified, that a validation of the model based on clinical data is possible only to a limited extent, or that model parameters are unknown or uncertain. For this model, the analysis of growth patterns of recurrent tumors may provide some indication for the plausibility of the model for individual patients. However, further research on model validation based on follow-up imaging, including MRI, MR Spectroscopy, and PET is needed. Nevertheless, we argue that an application of the model may be helpful in radiotherapy target volume definition despite these limitations. The model parameterizes features of glioma growth that are commonly agreed upon. The model can be viewed as a component in an automatic GTV to CTV expansion tool, which defines a distance measure in the brain that accounts for the patient specific anatomy. It allows treatment planners to incorporate their belief on glioma growth characteristics in multiple stages: In the first stage, anatomical boundaries can be accounted for, which is achieved by segmentation of the ventricles, falx cerebri and tentorium cerebelli. This is a promising application because, clearly, the falx represents a barrier for migrating tumor cells while the corpus callosum allows for inter-hemispheric spread of disease. In the second stage, reduced gray matter infiltration can be incorporated by choosing $D_w/D_g > 1$. If integrated into contouring software used in clinical practice, this tool can suggest an initial target volume to the physician, which optionally can be modified manually.\\

\section{Conclusions}

The Fisher-Kolmogorov equation represents a model for the spatial distribution of infiltrating tumor cells in normal appearing brain tissue. It can be applied in radiotherapy target delineation by defining target volumes as isolines of the tumor cell density. The model can incorporate spatial growth characteristics of infiltrating gliomas: the effect of anatomical barriers and a reduced tumor infiltration in gray matter. Therefore, the tumor growth model can provide the basis for an automatic contouring tool that provides the option to account for widely agreed growth patterns of glioma. The approach appears particularly useful for tumors located close to the falx and the corpus callosum. The most crucial input to the model is the segmentation of the brain tissue, in particular the reliable segmentation of the anatomical barriers falx cerebri and tentorium cerebelli.


\appendix

\section{Target volume comparison for different tumor locations}
\label{SecAppendix}
In section \ref{SecOtherCases}, we summarized the results for model based target delineation for 10 GBM patients that were analyzed.  In this appendix, we provide details on four representative cases, which span a range of different tumor locations, i.e. GBMs in the parietal lobe, temporal lobe, frontal lobe, and bilateral corpus callosum. The Dice coefficients between the manual and model-derived target volumes are summarized in table \ref{TabDice}, together with the corresponding Dice coefficient between the 95\% isodose lines of the IMRT plans.

\begin{figure}[hbt]
\centering
\subfigure[]{
\includegraphics[height=4.95cm]{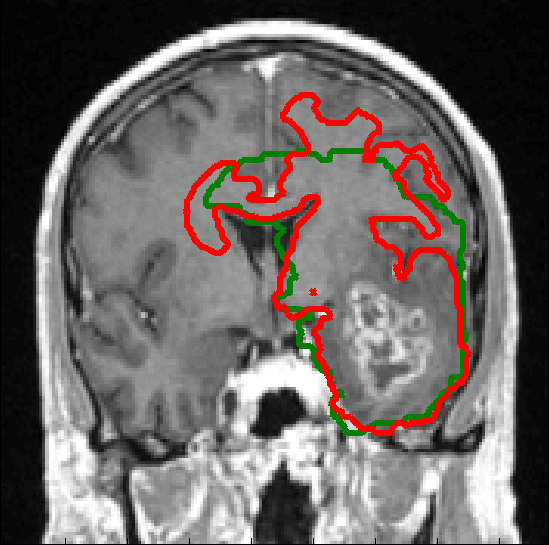}
}
\subfigure[]{
\includegraphics[height=5cm]{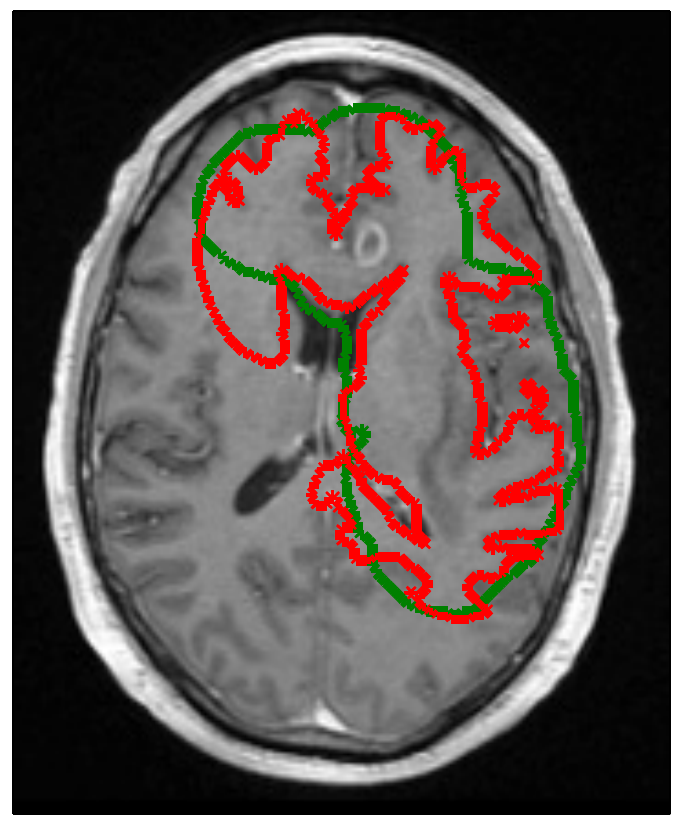}
}
\caption{Multifocal GBM located in the left temporal and frontal lobe: (a) coronal T1 post contrast image; (b) axial T1 post contrast image. The manually drawn CTV is shown in green, the model derived CTV in red.}
\label{FigPat003}
\end{figure}

\begin{figure}[hbtp]
\centering
\subfigure[cell density]{
\includegraphics[height=6cm]{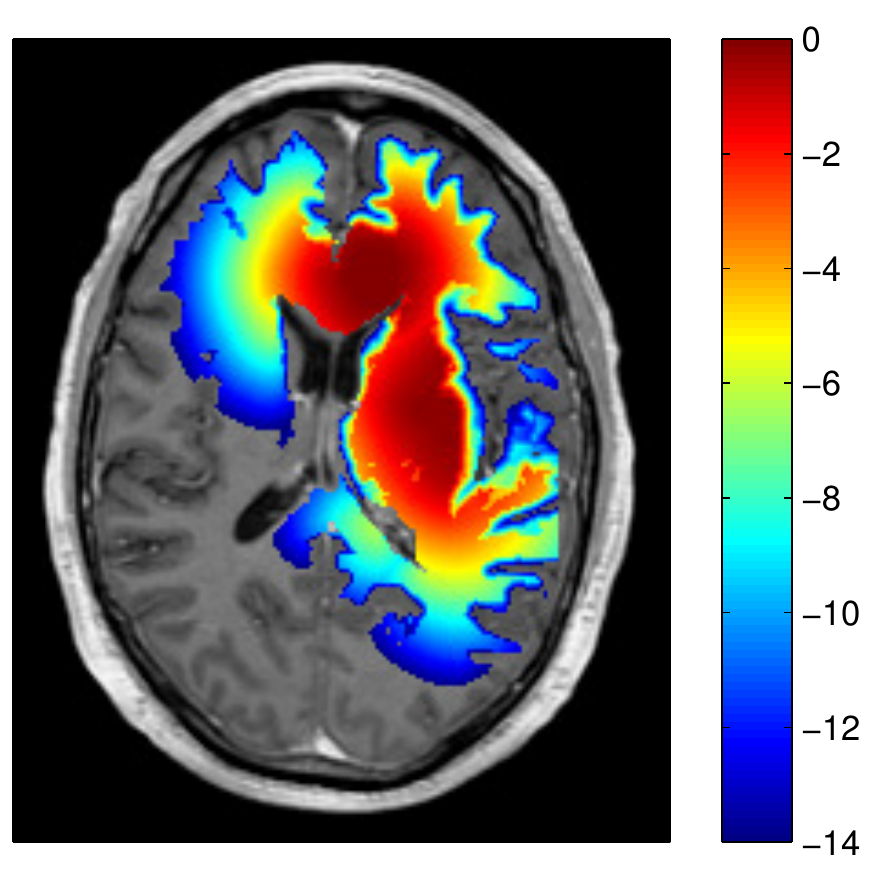}
}
\subfigure[model-derived target]{
\includegraphics[height=6cm]{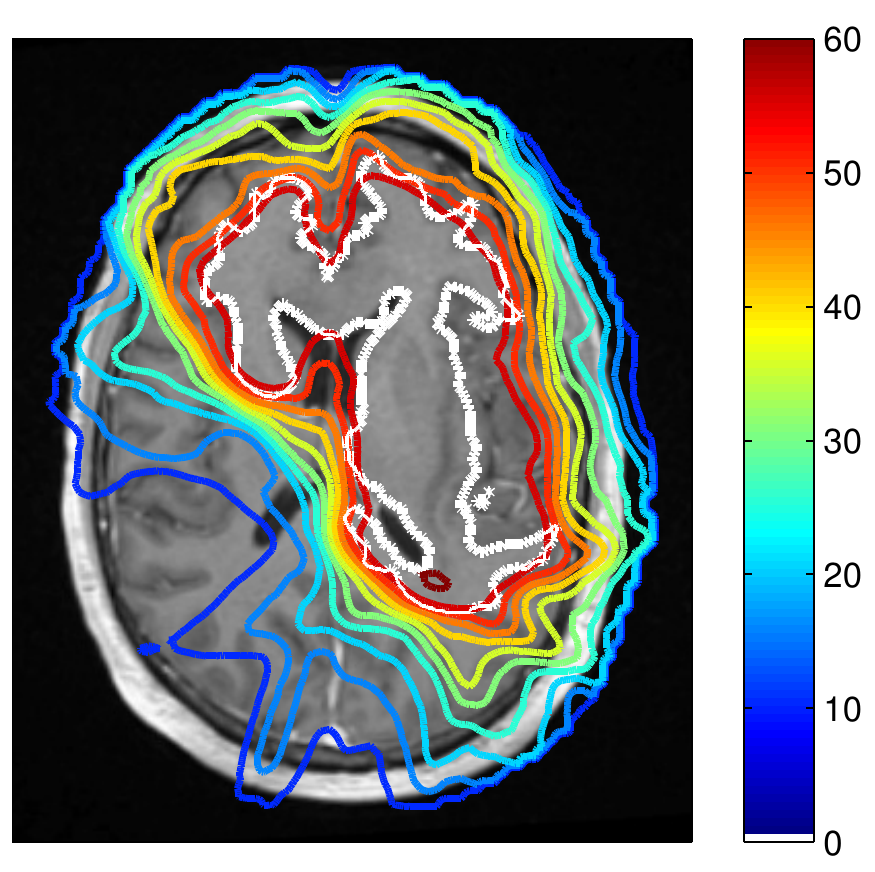}
}\\
\subfigure[manual target]{
\includegraphics[height=6cm]{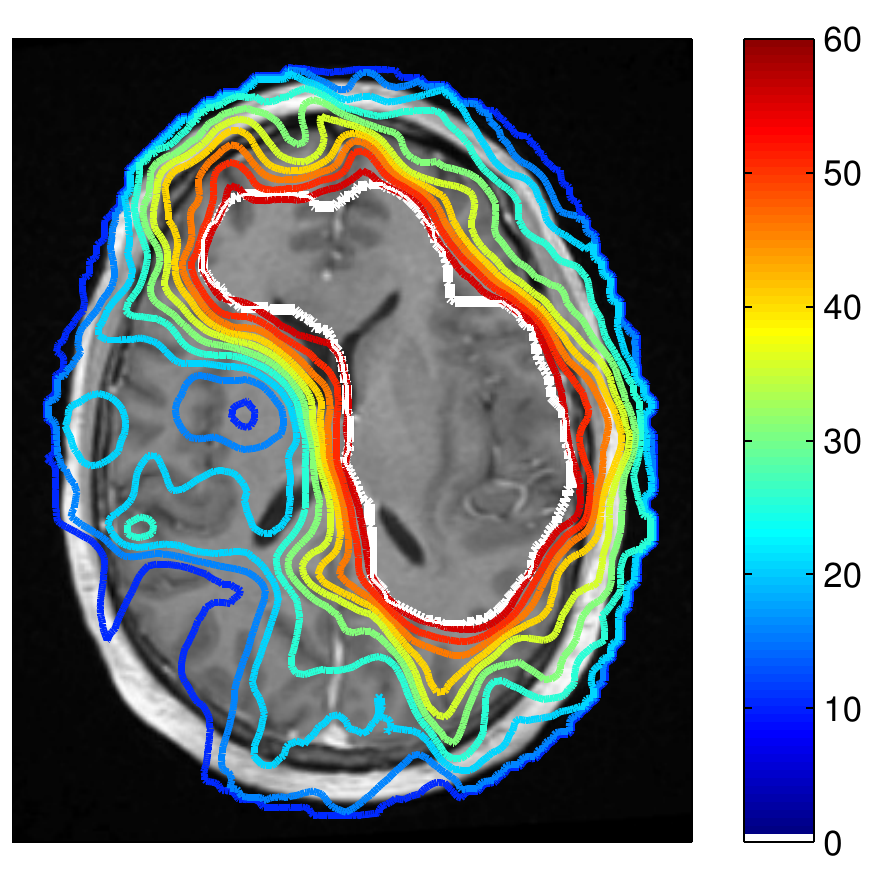}
}
\subfigure[dose difference]{
\includegraphics[height=6cm]{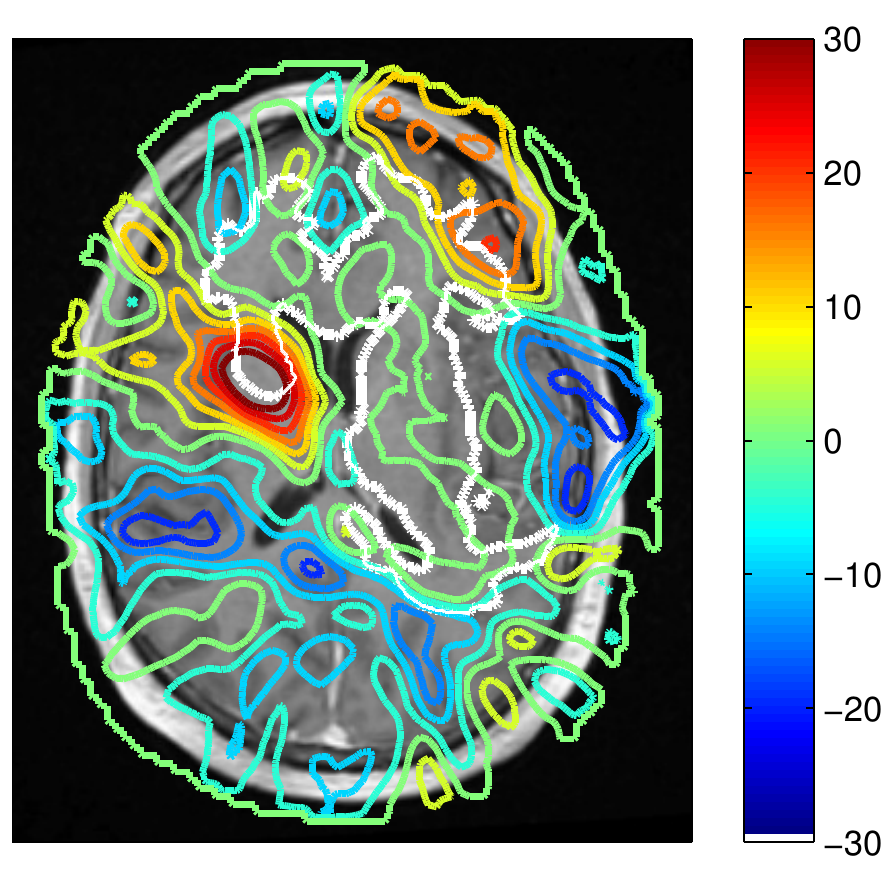}
}
\caption{(a) Simulated tumor cell density for the patient shown in figure \ref{FigPat003}; (b-c) Comparison of IMRT dose distributions (in units of Gy) of the plan based on manual target volume (c) and the model-derived target volume (b). The difference of the dose distributions is shown in (d).}
\label{FigPat003Dose}
\end{figure}

\subsection{Multifocal temporal/frontal lobe case}

Figure \ref{FigPat003} shows a multifocal GBM case involving the left temporal lobe as well as the frontal lobe. Figure \ref{FigPat003}a shows the coronal T1 post contrast image, revealing the contrast enhancing tumor mass in the temporal lobe. The axial T1 post contrast image in figure \ref{FigPat003}b shows the lesion in the frontal lobe. The simulated tumor cell density in figure \ref{FigPat003Dose}a  illustrates several features discussed in sections \ref{SecCase} and \ref{SecPlanning}: The tumor growth model describes the steep fall-off of the tumor cell density in gray matter, leading to differences in the target volume around the lateral sulcus (figure \ref{FigPat003}b). In addition, modeling tumor cell infiltration through the corpus callosum leads to differences in the target volume in the contralateral frontal lobe. Figure \ref{FigPat003Dose} shows the IMRT plan comparison for a homogeneous 60 Gy prescription to the manual CTV (c) and the model-derived CTV (b). The figure shows that the differences in the target volume partly translate into differences in the dose distribution. In particular, the dose difference plot in figure \ref{FigPat003Dose}d shows a lower dose in the lateral sulcus region for the model-based plan, and a higher dose in the contralateral hemisphere close to the corpus callosum.

\subsection{Parietal lobe case}
\begin{figure}[hbt]
\centering
\subfigure[]{
\includegraphics[height=5cm]{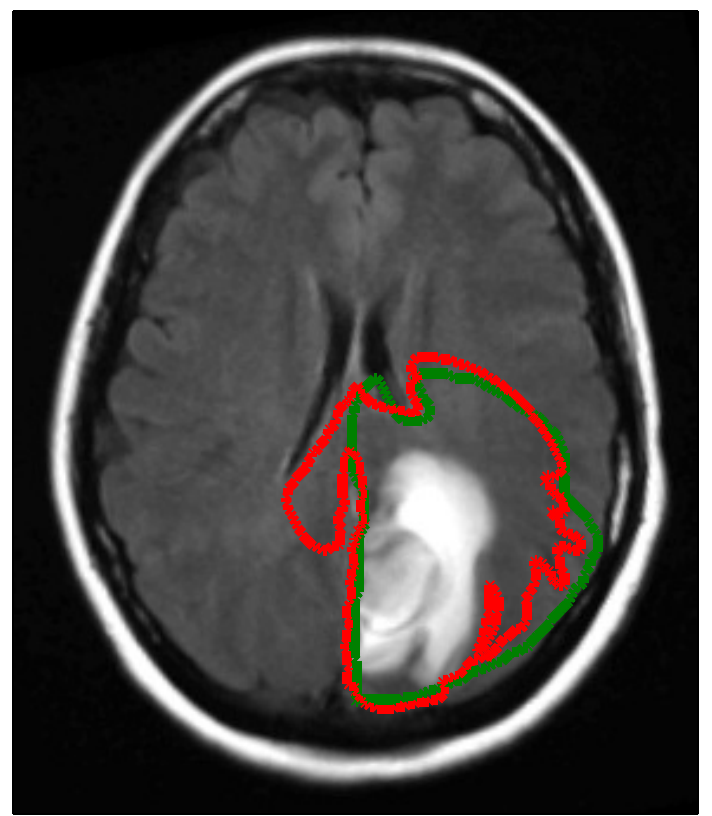}
}
\subfigure[]{
\includegraphics[height=5cm]{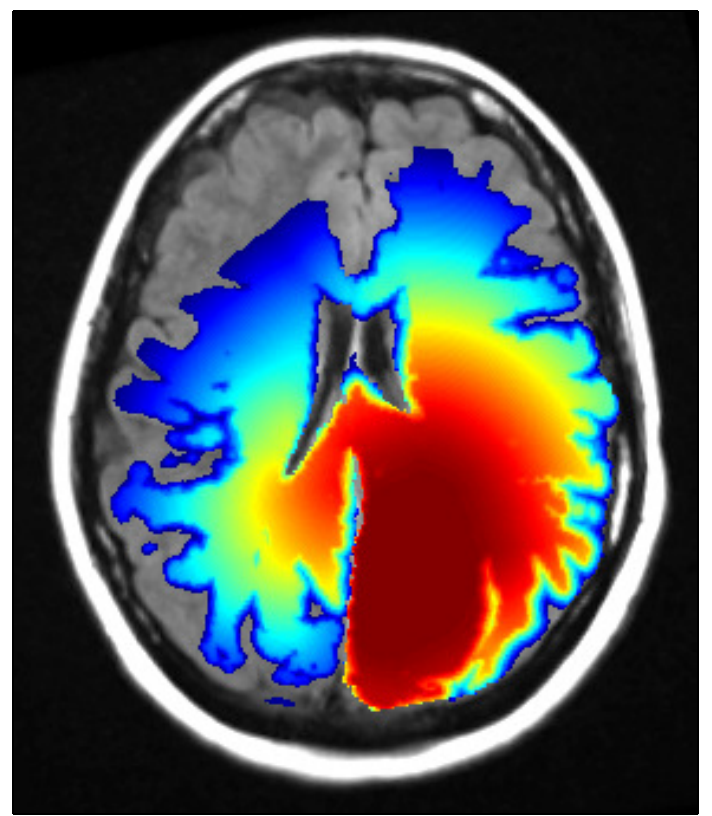}
}
\caption{GBM located in the left parietal lobe adjacent to the falx: (a) T2-FLAIR image with manual CTV (green) and model derived CTV (red); (b) simulated tumor cell density.}
\label{FigPat004}
\end{figure}

Figure \ref{FigPat004}a shows the T2-FLAIR image of a GBM located in the left parietal lobe. Compared to the case discussed in section \ref{SecCase}, the tumor is located more posteriorly and closer to the falx. The simulated tumor cell density for $\dwdg = 100$ is shown in figure \ref{FigPat004}b. Similar to the case discussed in section \ref{SecCase}, the tumor growth model simulates the infiltration of the contralateral side via the corpus callosum. The comparison of model derived target volume (red) to the manually drawn CTV (green) in figure \ref{FigPat004}a reveals that the tumor growth model suggests further spread into the contralateral hemisphere.

\subsection{Temporal lobe case}

\begin{figure}[hbt]
\centering
\subfigure[]{
\includegraphics[height=5cm]{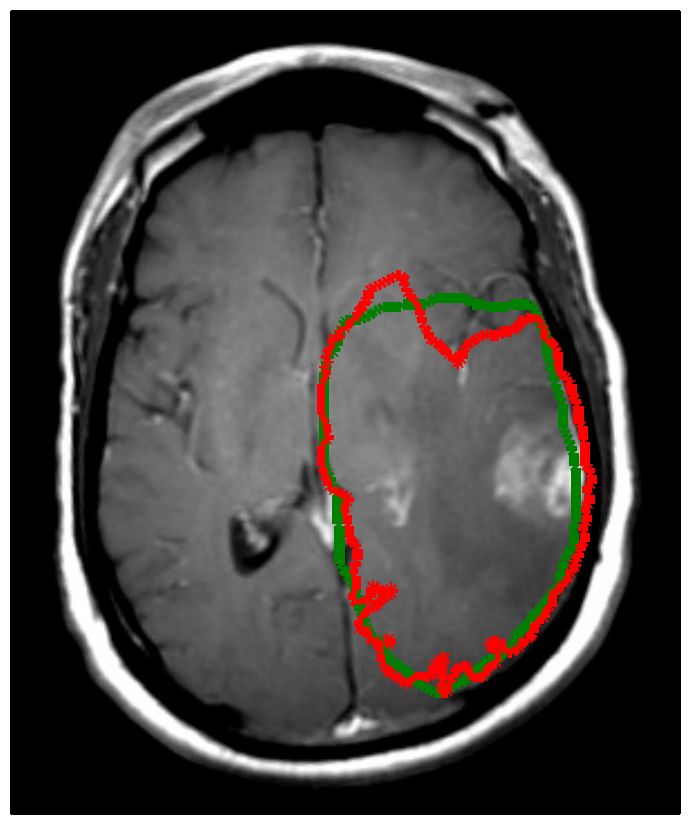}
}
\subfigure[]{
\includegraphics[height=5cm]{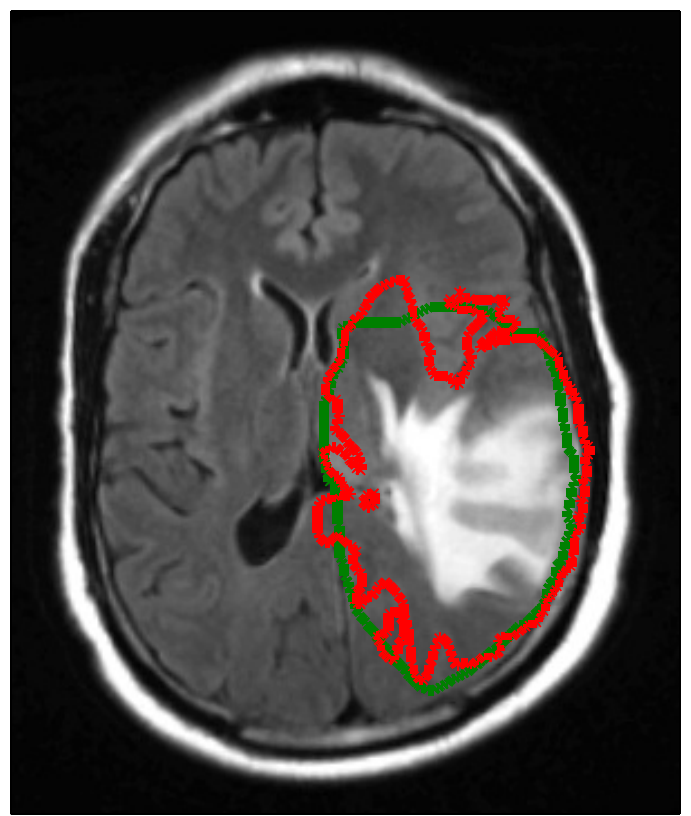}
}
\subfigure[]{
\includegraphics[height=5cm]{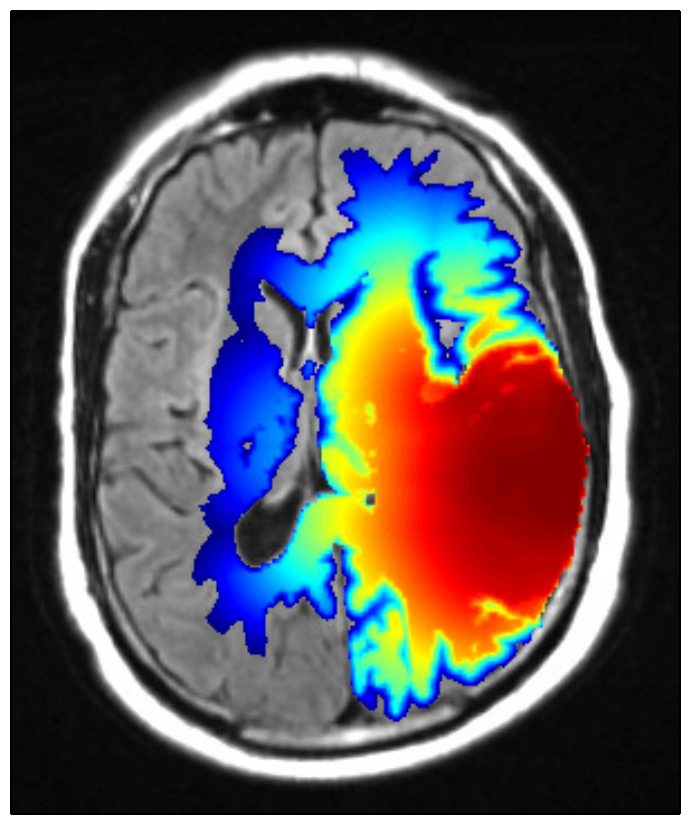}
}
\caption{GBM located in the left temporal lobe: (a) T1 post contrast image; (b) T2-FLAIR image; (c) simulated tumor cell density. Note that (b) and (c) represent the same slice whereas (a) is located 7.5 mm inferiorly. The manually drawn CTV is shown in green, the model derived CTV in red.}
\label{FigPat006}
\end{figure}

Figure \ref{FigPat006} shows a GBM located in the left temporal lobe. As for the other patients, the ventricles and the falx represent a boundary for infiltrating tumor cells, whereas the corpus callosum allows for contralateral spread. However, in this case, the tumor is located relatively far from the corpus callosum. Therefore, the model based target contour does not differ significantly from the manually drawn CTV in the region around the midsagittal plane (figure \ref{FigPat006}b). The largest differences between manual and model based CTV occur in the regions near the lateral sulcus, where the tumor growth model describes the sulcus with its surrounding gray matter as a boundary (figures \ref{FigPat006}a/b). Note that the left lateral sulcus is shifted anteriorly due to the tumor with mass effect.

\subsection{Corpus callosum case}

\begin{figure}[hbt]
\centering
\subfigure[]{
\includegraphics[height=5cm]{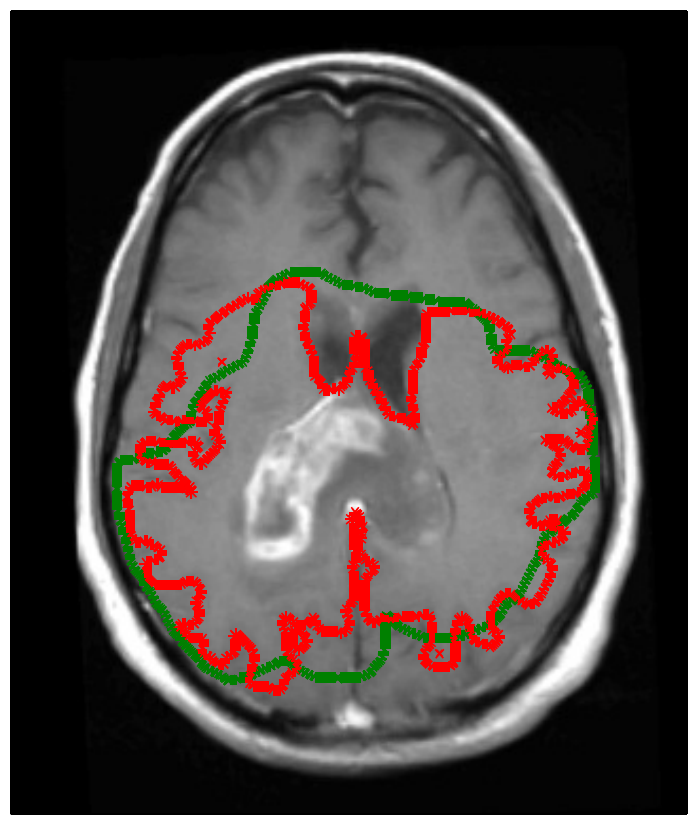}
}
\subfigure[]{
\includegraphics[height=5cm]{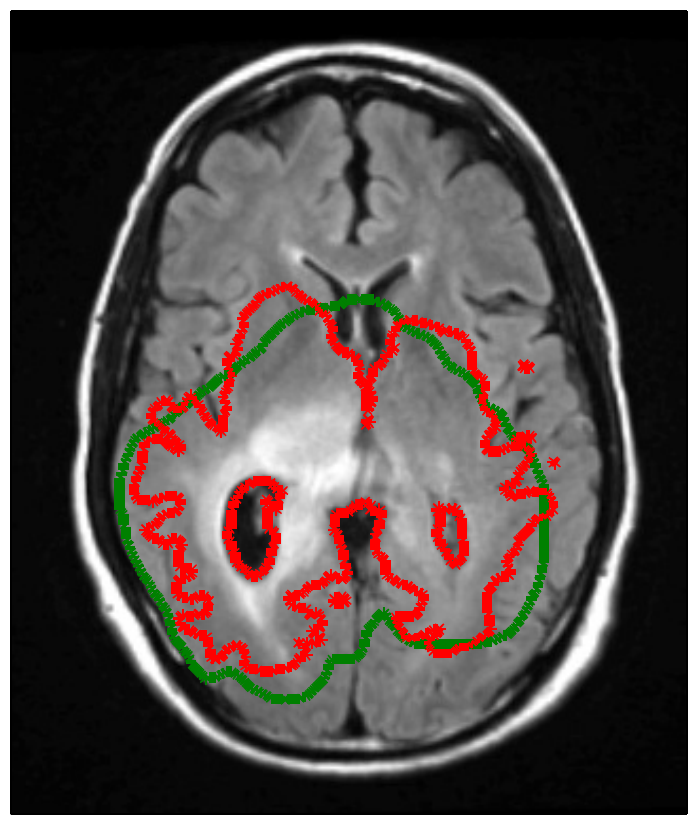}
}
\subfigure[]{
\includegraphics[height=5cm]{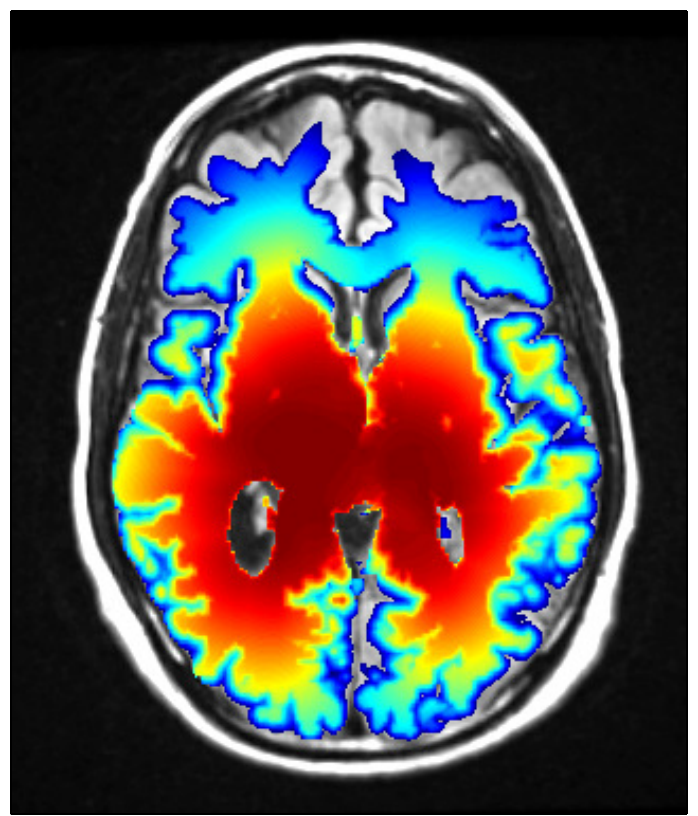}
}
\caption{GBM located in the corpus callosum: (a) T1 post contrast image; (b) T2-FLAIR image; (c) simulated tumor cell density. Note that (b) and (c) represent the same slice whereas (a) is located 12.5 mm superiorly. The manually drawn CTV is shown in green, the model derived CTV in red.}
\label{FigPat015}
\end{figure}

Figure \ref{FigPat015} shows a GBM case involving the bilateral corpus callosum.  The simulated tumor cell density in \ref{FigPat015}c models the fall-off of the tumor cell density in the cortical gray matter surrounding the sulci. This is most prominent in the proximity of major sulci including the lateral sulcus and occipital lobe. The effect of reduced gray matter infiltration leads to differences of up to approximately 1 cm between manual and model derived CTV. In regions where the CTV extends all the way to the dura (see e.g. right posterior region in figure \ref{FigPat015}a), the differences are insignificant because of the limited dose gradients that can be achieved with therapeutic photon beams. However, in several regions, differences in the CTV contours translate into dose differences in an IMRT plan. This is visible in figure \ref{FigPat015}b, where the red contour is further expanded into the major fiber tracts of the corona radiata, but the ventricles and regions in the occipital lobe are spared.

\section*{Acknowledgment}
The project was supported by the Federal Share of program income earned by Massachusetts General Hospital on C06 CA059267, Proton Therapy Research and Treatment Center. Additional support was provided by the ERC Advanced Grant MedYMA.



\end{document}